\documentclass[letterpaper,prl,aps,twocolumn,showpacs,superscriptaddress,nofootinbib]{revtex4-1}
\usepackage{bm,color}
\usepackage{mathptmx}
\usepackage[T1]{fontenc}
\usepackage[utf8]{inputenc}
\usepackage{latexsym}
\usepackage{graphicx} %includes pictures in the text
\usepackage{amsmath} %better view for mathematical operations
\usepackage{amsfonts} %makes set symbols available through \mathbb{}
\usepackage{braket}
\usepackage{subfigure}
\usepackage{lineno}

\def \be{{\bf e}}

	\newcommand{\bbm}{\begin{pmatrix}}
	\newcommand{\ebm}{\end{pmatrix}}

	\usepackage[normalem]{ulem} %makes underlining possible: \uline \uwave \sout
	\definecolor{mgreen}{RGB}{1,123,0}

\def\be{\begin{equation}}
\def\ee{\end{equation}}
\begin{document}
%\linenumbers

\title{Quantum gas magnifier for sub-lattice-resolved imaging of three-dimensional quantum systems}

\author{Luca Asteria}
\affiliation{Institut für Laserphysik, Universität Hamburg, 22761 Hamburg, Germany}
\author{Henrik P. Zahn}
\affiliation{Institut für Laserphysik, Universität Hamburg, 22761 Hamburg, Germany}
\author{Marcel N. Kosch}
\affiliation{Institut für Laserphysik, Universität Hamburg, 22761 Hamburg, Germany}
\author{Klaus Sengstock}
\email{klaus.sengstock@physnet.uni-hamburg.de}
\affiliation{Institut für Laserphysik, Universität Hamburg, 22761 Hamburg, Germany}
\affiliation{The Hamburg Centre for Ultrafast Imaging, 22761 Hamburg, Germany}
\affiliation{Zentrum für Optische Quantentechnologien, Universität Hamburg, 22761 Hamburg, Germany}
\author{Christof Weitenberg}
\affiliation{Institut für Laserphysik, Universität Hamburg, 22761 Hamburg, Germany}
\affiliation{The Hamburg Centre for Ultrafast Imaging, 22761 Hamburg, Germany}

\maketitle

{\bf Imaging is central for gaining microscopic insight into physical systems, but direct imaging of ultracold atoms in optical lattices as modern quantum simulation platform suffers from the diffraction limit as well as high optical density and small depth of focus. We introduce a novel approach to imaging of quantum many-body systems using matter wave optics to magnify the density distribution prior to optical imaging, allowing sub-lattice spacing resolution in three-dimensional systems. Combining the site-resolved imaging with magnetic resonance techniques for local addressing of individual lattice sites, we demonstrate full accessibility to local information and local manipulation in three-dimensional optical lattice systems. The method opens the path for spatially resolved studies of new quantum many-body regimes including exotic lattice geometries.}

%%%%%%%%%%%%%%%%%%%%%%%%%%%%%%%%%%%%%%%
%\section{Introduction}
%%%%%%%%%%%%%%%%%%%%%%%%%%%%%%%%%%%%%%%

Experimentally driven understanding of quantum mechanical phenomena depends crucially on the possibility to observe them at the microscopic level. The quantum nature of matter shows itself on small scales, which has triggered tremendous efforts to develop advanced methods with increasing resolution to image the quantum system itself. Here, we introduce the alternative approach based on the idea to first magnify the quantum system itself to more accessible scales, which can then be easily imaged. This significantly reduces limitations such as finite resolution, optical density, and most importantly depth of focus, thus allowing to image three-dimensional (3D) systems. We demonstrate this approach in a quantum simulator composed of quantum gases in the form of ultracold atoms in optical lattices and realize imaging of 3D systems with sub-lattice resolution. 

	\begin{figure}[h]
		\includegraphics[width=0.85\linewidth]{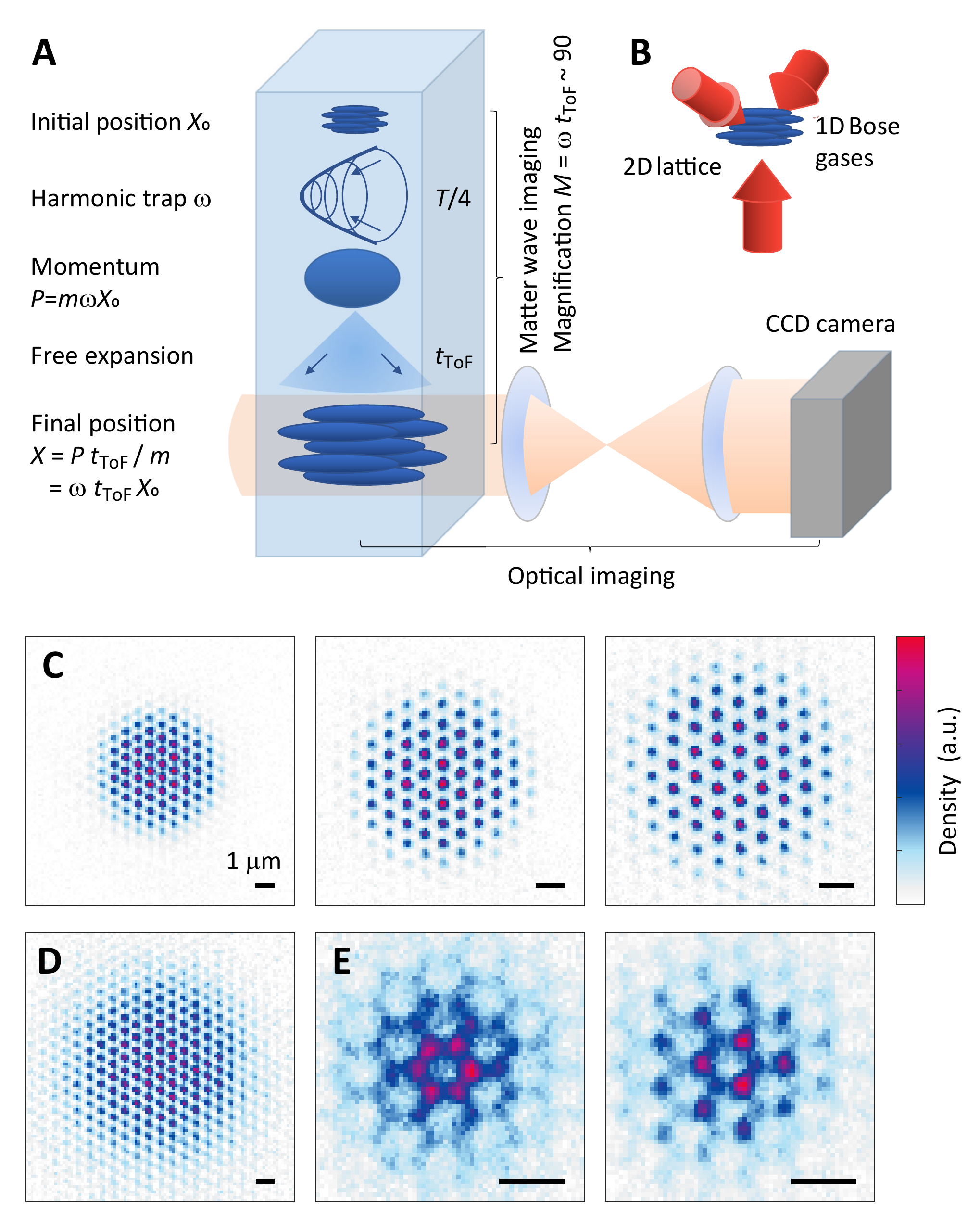}
		\caption{{\bf Working principle of the quantum gas magnifier using matter wave optics}. ({\bf A}) The density distribution of ultracold atoms in an optical lattice is magnified by matter wave optics composed of a pulsed dynamics in a harmonic trap and a free expansion. Subsequently, it can be imaged with optical absorption imaging of moderate resolution and without restrictions from optical density or depth of focus. ({\bf B}) Sketch of the 2D hexagonal optical lattice. ({\bf C}) Images of ultracold bosonic atoms in a 2D triangular lattice for constant system size given by the confinement $\omega_{\rm system}/2\pi=225\,$Hz, but varying magnification of $M=43(1)$, $65(1)$, $80(1)$ (from left) tuned via tighter magnetic confinements $\omega_{\rm pulse}$ during the matter wave optics. ({\bf D}) Image of a larger system with confinement $\omega_{\rm system}/2\pi=89\,$Hz imaged with magnification $M=43(1)$. ({\bf E}) Images of a honeycomb lattice and a boron nitride lattice with a sublattice offset of $4.6$~kHz with a magnification of $M=89(1)$. The scale bars have a length of 1\,µm. The atom number is in between 48,000 and 59,000 for the six images.}\label{fig:1_Idea}
		\end{figure} 

Direct optically resolved imaging of ultracold atoms in optical lattices, known as quantum gas microscopy \cite{Bakr2010,Sherson2010}, requires very high numerical apertures and is so far restricted to two-dimensional (2D) systems due to the fundamental limitation of the depth of focus and to unit lattice site occupation due to light-assisted collisions. The depth of focus can be overcome by using an electron microscope \cite{Gericke2008} or an ion microscope \cite{Veit2021}, but at the cost of a reduced detection efficiency and a large technological complexity. Recent experiments have reached sub-lattice resolution via superresolution microscopy using non-linear atom-light interactions \cite{McDonald2019, Subhankar2019}, but relying on scanning techniques. Our quantum gas magnifier does not suffer from these limitations and extends sub-lattice-site resolved imaging to new 3D regimes such as bosons or fermions in 3D optical lattices or sub-wavelength lattices with drastically enhanced energy scales \cite{Ritt2006, Nascimbene2015, Wang2018, Anderson2020}. The technique yields full single-shot images, which gives direct access to correlations and, e.g., spontaneous pattern formation. Furthermore, the concept can be applied and adapted to very different physical systems such as exotic atomic species or mixtures. 

Our quantum gas magnifier uses matter wave optics in the time domain to magnify the atomic density distribution before the standard optical absorption imaging. To this end, a harmonic potential of trapping frequency $\omega_{\rm pulse}=2\pi/T$ is applied for a time $T/4$ mapping the spatial distribution to the momentum distribution \cite{Shvarchuck2002, VanAmerongen2008, Tung2010, Murthy2014}. This is initialized in our case by switching off the lattice. This matter wave lens is followed by free time-of-flight expansion (ToF) of duration $t_{\rm ToF}$. This combination reproduces the initial spatial distribution with a magnification $M\approx \omega_{\rm pulse}\cdot t_{\rm ToF}$ (Fig.\,\ref{fig:1_Idea}A). Note that also more complex pulsed lenses and other time-domain optical elements can be used in this scheme. An advantage of combining a $T/4$ pulse with time of flight is that the aberrations introduced by the finite ToF can be perfectly compensated by choosing the evolution time in the harmonic trap slightly above $T/4$ ~\cite{SupMat}.

Fig.\,\ref{fig:1_Idea}C-E demonstrates the power of this method with the first single-shot site-resolved images of a 3D quantum gases in 2D optical lattices including images of lattices with two-atomic basis. In the following, after describing the concept more closely, we additionally demonstrate high resolution thermometry across the thermal-to-BEC phase transition for a 3D quantum gas in a triangular optical lattice as well as full local addressability and precision measurements of thermally activated dynamics in a lattice system. Finally, we also demonstrate sub-wavelength resolution to study local dynamics. The flexibility and adaptability of our concept now allow for very precise locally resolved and locally controlled measurements of higher-dimensional quantum gas systems.

%%%%%%%%%%%%%%%%%%%%%%%%%%%%%%%%%%%%%%%%%%%%%%%%%%%%%%%%%%%
%\section{Working principle of the quantum gas magnifier}
%%%%%%%%%%%%%%%%%%%%%%%%%%%%%%%%%%%%%%%%%%%%%%%%%%%%%%%%%%%

The experiments presented here use $^{87}$Rb Bose-Einstein condensates (BEC) evaporatively cooled in a magnetic trap. The potential of the magnetic trap is in-plane radially symmetric with trapping frequency which is ramped within 100\,ms to $\omega_{\rm system}=2\pi\cdot[89-658]\,$Hz. We ramp up triangular or honeycomb optical lattices formed by the interference of lattice beams of wavelength $\lambda=1064\,{\rm nm}$ leading to a lattice constant of $a_{\rm lat}=2\lambda/3=709\,{\rm nm}$, which sets the energy scale $E_{\rm rec}=h^2/(2 m \lambda^2)$ for the lattice depth, where $h$ is Planck's constant and $m$ the atomic mass. The harmonic transverse confinement has a trapping frequency $\omega_z$ of typically $2\pi \times 29$~Hz, resulting in a Josephson junction array of BECs in the tubes of the 2D lattice. The trap frequency is then ramped to $\omega_{\rm pulse}$ for the magnification protocol. The magnetic trap is suitable for the $T/4$ evolution because of its smoothness, radial symmetry and strong confinement: For typical parameters of $t_{\rm ToF}\approx 20\,$ms and $\omega_{\rm pulse}/(2\pi)$ up to $\approx 700\,$Hz we measure large magnifications of up to $M=93(1)$, allowing resolution of the lattice spacing with conventional absorption imaging with magnification 2 on a CCD camera (Fig.\,\ref{fig:1_Idea}). 

The resolution of the quantum gas magnifier can be made very high because the harmonic trap has a large spatial extension corresponding to a large numerical aperture of the matter wave optics. In practice the resolution is mainly limited by the convolution with our optical imaging resolution \cite{SupMat}. The effect of interactions during the magnification protocol can be effectively suppressed by working with incoherent systems or by removing the coherence via freezing in a deep lattice \cite{SupMat}. 

	\begin{figure}[t]
		\includegraphics[width=0.85\linewidth]{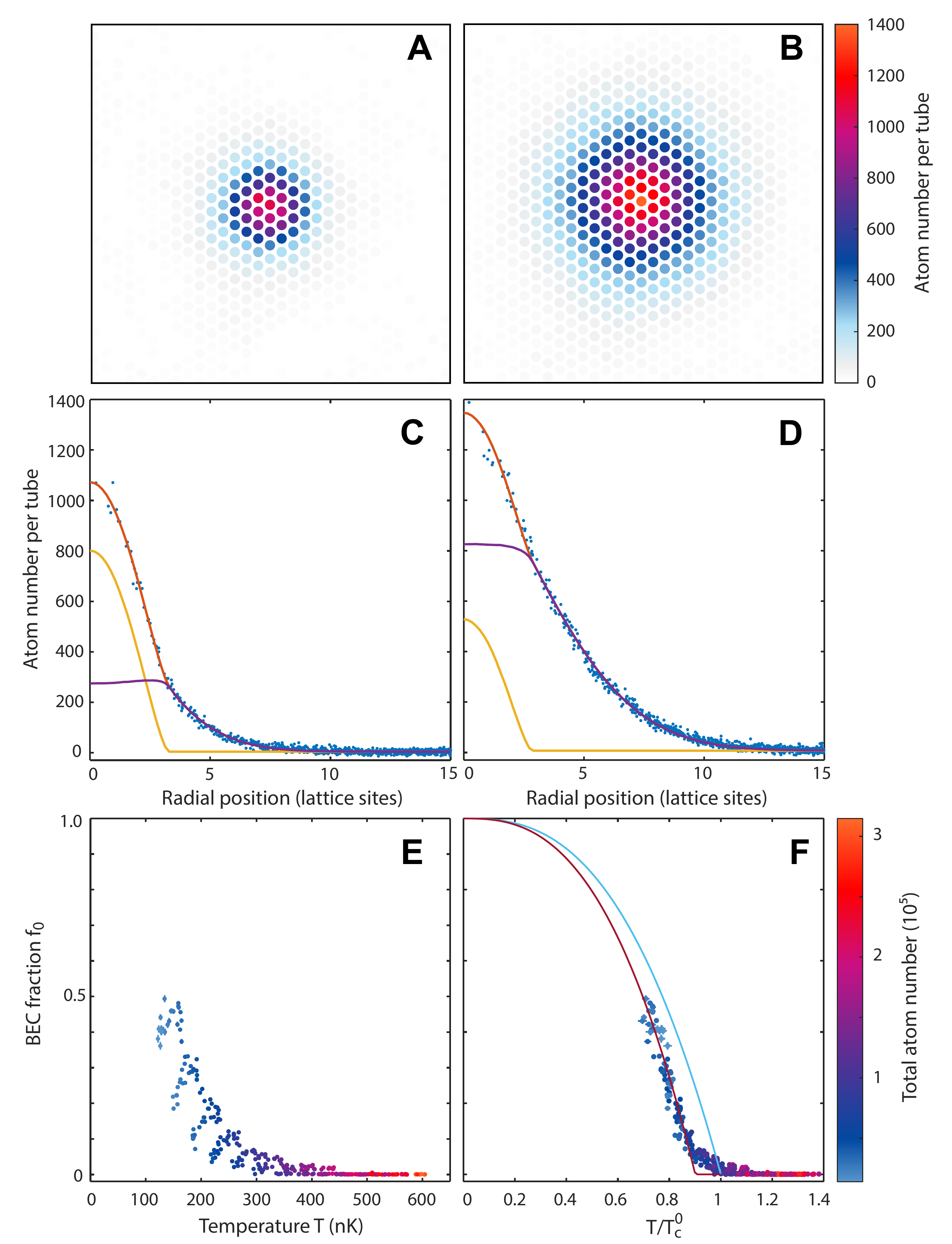}
		\caption{{\bf Thermal-to-BEC phase transition in optical lattices observed via high-resolution density profiles}. ({\bf A} and {\bf B}) Spatial density distributions of BECs in triangular optical lattices prepared at different temperatures and atom numbers of 171(1)~nK and 37,000(400) atoms (A) and 310(1)~nK and 106,000(600) atoms (B). The densities are shown as atom numbers per lattice site from integration over the Wigner-Seitz cells. ({\bf C} and {\bf D}) Radial density profile corresponding to (A) and (B), respectively, with a bimodal fit (orange line) consisting of the condensed part (yellow line) and the thermal part (purple line).  ({\bf E} and {\bf F}) Condensate fraction (circles) obtained from the bimodal fits as a function of the temperature (E) and of the temperature in units of the scaling temperature $T_{\rm c}^0$ (F). Most error bars are smaller than the symbol size.  The light blue line in (F) shows the power law approximation of the non-interacting theory described in the main text. The purple line is a fit to the data with the same power law. The bandwidth of the lowest band is $k_{\rm B} \cdot 5.4\,$nK and the gap between the first and second band is $k_{\rm B} \cdot 290$ nK. The color encodes the total atom number of the clouds. All error bars correspond to the $68\%$ confidence interval.}\label{fig:2_profile}
		\end{figure} 

%%%%%%%%%%%%%%%%%%%%%%%%%%%%%%%%%%%%%%%%%%%%%%%
%\section{High resolution density profiles}
%%%%%%%%%%%%%%%%%%%%%%%%%%%%%%%%%%%%%%%%%%%%%%%

As a first benchmark experiment, we study the thermal-to-BEC phase transition in a lattice of tubes allowing us to confirm the faithful imaging of lattice site occupations. Furthermore, we show how the high-resolution access to real space profiles via the magnifier provides an excellent approach to optical lattice thermometry, which is notoriously difficult from the more common momentum space images \cite{McKay2009, Trotzky2010, Cayla2018}.

In order to study the phase transition, we prepare the system at varying temperature and atom number by adjusting the end point of the evaporation ramp and a varying hold time before the lattice loading. For the analysis, we start with the extraction of the on-site populations (Fig.~\ref{fig:2_profile}A,B) and determine the high-precision radial density profiles making use of the well characterized radially symmetric trap (Fig.~\ref{fig:2_profile}C,D). The data can be described by a bimodal model consisting of a condensed part and a thermal part including the repulsion of the thermal atoms from the condensate in mean-field approximation~\cite{SupMat}. The model is fitted to the 2D distribution and the excellent fit quality of the radial profiles (Fig.~\ref{fig:2_profile}C,D) confirms the exact measurement of the lattice site occupations.

The fit allows us to extract the temperature $T$ from the thermal component and the condensate fraction $f_0$ from the atom numbers in the two components with very high precision. Owing to the dependence of the critical temperature $T_{\rm c}$ on the total atom number, the condensate fraction as a function of temperature does not result in a single curve (Fig.~\ref{fig:2_profile}E). In order to describe this dependence we set up an analytic non-interacting model predicting the critical temperature $T_{\rm c}^0$ to renormalize the experimental temperatures using $T_{\rm c}^0$ as a scaling temperature, resulting in a collapse of the data on a single curve (Fig.~\ref{fig:2_profile}F). We observe a shift of the critical temperature towards lower values compared to the non-interacting model. To quantify this shift we approximate the non-interacting model by a power law in the density of states resulting in a description $f_0 = (1 - T/T_{\rm c})^\alpha$ with $\alpha = 2.69(1)$ characterizing the underlying density of states interpolating between a lattice regime and a continuum regime~\cite{SupMat}. 

Fitting this function to the data satisfying $f_0 > 0.1$ results in $T_{\rm c} = 0.901(4)T_{\rm c}^0$, where the small statistical error reflects the excellent collapse on a single curve, thus showing the quality of the thermometry. Additionally, we estimate a systematic error of $1\%$ stemming from an uncertainty of the atom number calibration of $3\%$. A shift of this order of magnitude is expected from interactions and finite size~\cite{Giorgini1996}, but a closed theoretical model for our regime where both trap and lattice are relevant does not exist. With the enhanced interactions in the optical lattice, the shift is larger than those experimentally observed for BECs in a 3D harmonic traps for comparable atom numbers \cite{Ensher1996, Smith2011}. Interestingly, we observe a pronounced smoothing of the phase transition despite the rather large atom number, which might be due to the 2D-3D crossover geometry of an array of tubes. Our precision thermometry measurements thus provide a benchmark for future theoretical studies of phase transitions in such geometries. 

	\begin{figure}[t]
		\includegraphics[width=0.85\linewidth]{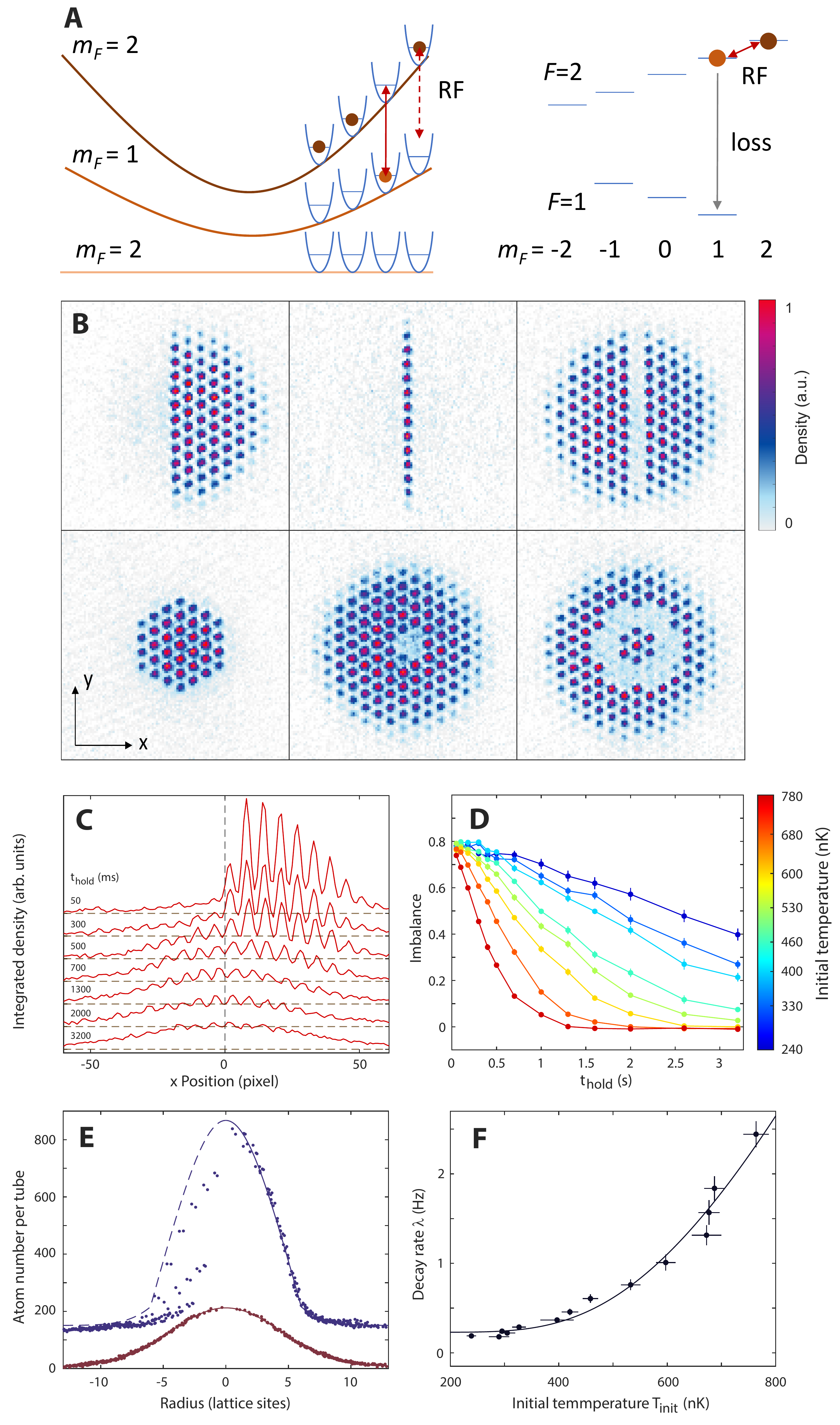}
		\caption{{\bf Local addressing and thermalization dynamics}. ({\bf A}) Scheme of local addressing via RF transitions in a magnetic field gradient and a sketch of the hyperfine states of $^{87}$Rb with the utilized RF transition and the loss channel. ({\bf B}) Example images of prepared density distributions. ({\bf C}) Single shot density profiles integrated along the $y$ direction for different hold times after removing the left half of the cloud illustrating the thermalization dynamics. The different profiles are offset for clarity. The initial temperature is $T_{\rm init}=0.76(2)$~$\mu$K. ({\bf D}) Time evolution of the imbalance for different initial temperatures. ({\bf E}) Radial density profiles after 50 ms (blue), shifted upwards for clarity, and 3.2 s (red) hold time with bimodal fits using only lattice site populations with positive $x$ positions larger than the maximally populated line (blue) and all populations (red) respectively. The data is averaged over 27 images (including the top and bottom data from (C)). The fit yields temperatures of 0.68(5)~$\mu$K and 1.25(4)~$\mu$K and demonstrates the reached thermal equilibrium. ({\bf F}) Dependence of the decay time on the initial temperature modelled by an Arrhenius process for thermal hopping and an offset rate for quantum tunneling (see text and~\cite{SupMat}).}\label{fig:4_addressing}
		\end{figure} 
		
%%%%%%%%%%%%%%%%%%%%%%%%%%%%%
%\section{Local addressing}
%%%%%%%%%%%%%%%%%%%%%%%%%%%%%

In a second set of experiments, we employ magnetic resonance (MR) techniques to realize local addressing of individual lattice sites and thereby demonstrate the full functionality of quantum gas microscopes without the need for large optical access thus making it compatible with other experimental constraints. While site-resolved addressing was previously also realized optically \cite{Weitenberg2011, Preiss2015} and with an electron beam \cite{Wuertz2009}, MR techniques are optimally suited for 3D systems and have, e.g., been proposed for wave-function engineering~\cite{Wigley2017}.

In the experimental protocol, we freeze the atomic distribution in a deep lattice and shift the magnetic trap ($\omega_{\rm addressing}/2\pi=543\,$Hz) by up to 20\,µm creating magnetic gradients between 23 and 50~kHz/µm at the atom's position. The magnetic gradient spatially splits the radio frequency (RF) transition between the initial stretched $F=2$, $m_F=2$ state and the target $F=2$, $m_F=1$ state and we drive spin flips at positions controlled via RF sweeps (Fig.\,\ref{fig:4_addressing}A). In order to empty the addressed lattice sites, we make use of the strongly spin-dependent loss rates driven by hyperfine-changing collisions, which are suppressed for the stretched initial spin state but empty the addressed lattice sites during the sweep time of 100 to 400~ms \cite{Schmaljohann2004}. When choosing $F$-changing transitions instead, the removal of one state could be achieved via an optical push out. The magnifier approach can also be easily extended to spin dependent imaging~\cite{SupMat}. By choosing the appropriate RF sweeps, we create very well resolved arbitrary patterns such as single lines, half systems or rings of varying radius (Fig.\,\ref{fig:4_addressing}B). 

Subsequently, we probe the thermalization dynamics of a system cut in half \cite{Choi2016} (Fig.\,\ref{fig:4_addressing}C). We monitor the thermalization via the imbalance $\mathcal{I}=(N_R-N_L)/(N_R+N_L)$ defined as the relative difference of the atom numbers $N_R$ in the right half and $N_L$ in the left half of the trap. The imbalance $\mathcal{I}$ decays to zero (Fig.\,\ref{fig:4_addressing}C,D) and we determine the thermalization rate from an exponential fit. We verify that the profiles with no imbalance are indeed in thermal equilibrium (Fig.\,\ref{fig:4_addressing}E).

The thermalization rate as a function of the initial temperature is almost constant up to temperatures of about $350\,$nK and then increases steeply with temperature (Fig.\,\ref{fig:4_addressing}F). We model this by an Arrhenius law describing thermal hopping combined with an offset rate resulting from quantum tunneling~\cite{SupMat}. We obtain a potential barrier height of $V_B = k_{\rm B} \cdot 2.4(6)\,$µK in excellent agreement with the peak-to-peak lattice depth of $k_{\rm B}~\cdot~2.6\,$µK deduced from lattice depth calibration and an offset rate $\Gamma_0 = 0.23(8)\,$Hz related to the tunneling energy $J=h \cdot 0.1\,$Hz of the lowest band. These experiments demonstrate that the quantum gas magnifier allows very precise spatially resolved studies of thermalization dynamics in optical lattices in new parameter regimes, which could be extended to strongly-correlated regimes by adding a transverse lattice.

	\begin{figure}[t]
		\includegraphics[width=0.85\linewidth]{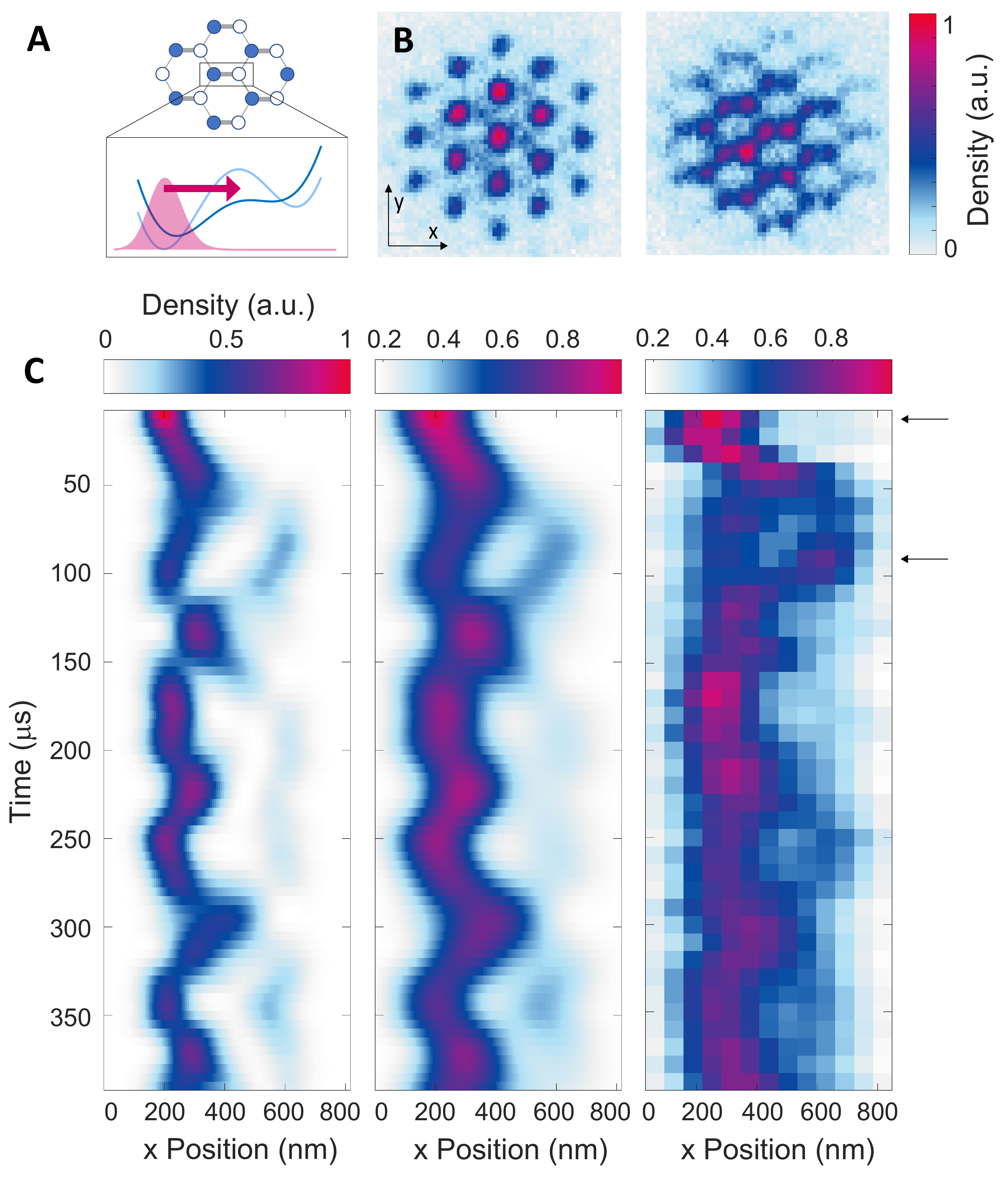}
		\caption{{\bf Nanoscale dynamics in a honeycomb optical lattice.} ({\bf A}) The honeycomb lattice with energy offset between the A sites (closed circles) and B sites (open circles) can be tuned into a lattice of dimers with stronger tunneling bonds along one direction (thicker grey lines). The inset shows cuts of the potential along a dimer before (light blue line) and after the quench (dark blue line) together with the initial density profile (red area). ({\bf B}) Experimental images for $10\,$µs, $90\,$µs after the quench. A lattice vector corresponds to $10.9$ pixel with a  magnification of $M=93(1)$. ({\bf C}) Time evolution of the density distributions within one dimer after the quench from the simulation (left) and from the experiment (right, cut of 1 pixel width). For a realistic comparison we have broadened the simulation results with a Gauss filter of $76\,$nm width and added an offset~\cite{SupMat} (middle). The arrows mark the evolution times shown in (B). 
	    }\label{fig:2_Quench}
		\end{figure}

%%%%%%%%%%%%%%%%%%%%%%%%%%%%%%%%%%
%\section{Nanoscale dynamics}
%%%%%%%%%%%%%%%%%%%%%%%%%%%%%%%%%%

Finally, we demonstrate the capability to resolve density features well below the lattice spacing by observing nanoscale dynamics after a quench of the lattice geometry. We start in a deep honeycomb lattice with large sublattice offset \cite{SupMat} leading to an initial population of the A-sublattice only and control the geometry by varying the imbalance of the lattice beam intensities $I_1$, $I_2$ and $I_3$. By abruptly reducing $I_2=I_3$ to $0.5\cdot I_1$, we create a lattice of dimers with enhanced tunnel coupling within the dimer as well as a displacement of the lattice sites (Fig.\,\ref{fig:2_Quench}A), thus exciting both a tunneling oscillation between the A and B sites and an oscillation within the lattice sites. 

The resulting dynamics of the atomic density within the dimer (averaged over all dimers with at least $50\%$ of the signal in the most populated dimer) is shown in (Fig.\,\ref{fig:2_Quench}B,C). We capture the dynamics by a non-interacting multi-band simulation including the finite switching time of the laser intensities of about 20~$\mu$s. The quantum gas magnifier on honeycomb optical lattices allows resolving the interplay of tunneling dynamics between lattice sites with nanoscale dynamics within the lattice sites \cite{McDonald2019, Subhankar2019} and opens a real-space approach to studying multi-orbital systems especially for extended 3D systems.

%%%%%%%%%%%%%%%%%%%%%%%%%%%%%%%%%
%\section{Outlook}
%%%%%%%%%%%%%%%%%%%%%%%%%%%%%%%%%

In conclusion, we have introduced a quantum gas magnifier based on matter wave optics and used it to image 3D quantum gases in triangular and honeycomb optical lattices with a resolution below the lattice spacing. Spatially resolved measurements give access to central scientific problems such as transport phenomena \cite{Choi2016}, spontaneous domain formation \cite{Parker2013}, or chiral edge and interface states in interacting topological matter \cite{Irsigler2019}. 
We estimate that the method can be pushed to a single-atom sensitive regime using free-space fluorescence imaging after the matter wave magnification, when the magnified lattice spacing is larger than the diffusive expansion from photon scattering \cite{Buecker2009,Bergschneider2018,SupMat} or using metastable helium and multi-channel plates \cite{Lawall1995}. This will allow for a direct study of correlations in strongly-interacting systems of atomic species, for which laser cooling and very deep optical lattices as in conventional quantum gas microscopes are not available. The magnification approach also circumvents pairwise atom loss during imaging in quantum gas microscopes, allowing measurements of many body systems with larger occupation number. 

Furthermore, the quantum gas magnifier can be employed to access coherence properties with high spatial resolution, e.g. by applying masks in Fourier space \cite{Murthy2019} or by magnification of interference phenomena like Talbot revivals \cite{Santra2017, SupMat}. We also expect that the sub-lattice spacing resolution would allow band resolved studies of multi-band systems.

{\bf Acknowledgments}. The work is funded by the Cluster of Excellence ’Advanced Imaging of Matter’ of the Deutsche Forschungsgemeinschaft (DFG) - EXC 2056 - project ID 390715994 and by the DFG Research Unit FOR 2414, project ID 277974659. C.W. acknowledges funding by the European Research Council (ERC) under the European Union’s Horizon 2020 research and innovation programme under grant agreement No. 802701.

{\bf Author contributions}. L.A., H.Z., and M.K. took and analyzed the experimental data and performed numerical simulations under the supervision of K.S. and C.W. All authors contributed to the interpretation of the results and to the writing of the manuscript.

%\newpage
%\appendix
%\section{SUPPLEMENTARY MATERIALS}

%add supplementary here

%\bibliography{mybib}


%\bibliographystyle{apsrev4-1}
%\bibliographystyle{naturemag}


\setcounter{equation}{0}
\setcounter{figure}{0}
\setcounter{table}{0}
%\setcounter{page}{1}
%\makeatletter
\renewcommand{\theequation}{S\arabic{equation}}
\renewcommand{\thefigure}{S\arabic{figure}}
\renewcommand{\bibnumfmt}[1]{[S#1]}
\renewcommand{\citenumfont}[1]{S#1}

%%%%%%%%%%%%%%%%%%%%%%%%%%%%%%%%%%%%%%%%%%%%%%%%%%%%%%%%%%%%%%5

\end{document}

%\linenumbers

\title{Supplementary Materials for:\\Quantum gas magnifier for sub-lattice-resolved imaging of three-dimensional quantum systems}

\author{Luca Asteria}
\affiliation{Institut für Laserphysik, Universität Hamburg, 22761 Hamburg, Germany}
\author{Henrik P. Zahn}
\affiliation{Institut für Laserphysik, Universität Hamburg, 22761 Hamburg, Germany}
\author{Marcel N. Kosch}
\affiliation{Institut für Laserphysik, Universität Hamburg, 22761 Hamburg, Germany}
\author{Klaus Sengstock}
\email{klaus.sengstock@physnet.uni-hamburg.de}
\affiliation{Institut für Laserphysik, Universität Hamburg, 22761 Hamburg, Germany}
\affiliation{The Hamburg Centre for Ultrafast Imaging, 22761 Hamburg, Germany}
\affiliation{Zentrum für Optische Quantentechnologien, Universität Hamburg, 22761 Hamburg, Germany}
\author{Christof Weitenberg}
\affiliation{Institut für Laserphysik, Universität Hamburg, 22761 Hamburg, Germany}
\affiliation{The Hamburg Centre for Ultrafast Imaging, 22761 Hamburg, Germany}

\maketitle

\tableofcontents

\section{Matter wave imaging theory}

\subsection{Focusing condition beyond the far-field limit}
In the following, we want to derive the focusing condition for the matter wave optics imaging composed of a quarter period evolution in a harmonic oscillator (HO) trap and a time-of-flight expansion. In particular, we prove that exact imaging can be obtained without having to reach the far field limit of the time-of-flight expansion by adjusting the evolution time in the harmonic trap. We give the derivation for the 1D case. 

\begin{figure}[t]
		\includegraphics[width=\columnwidth]{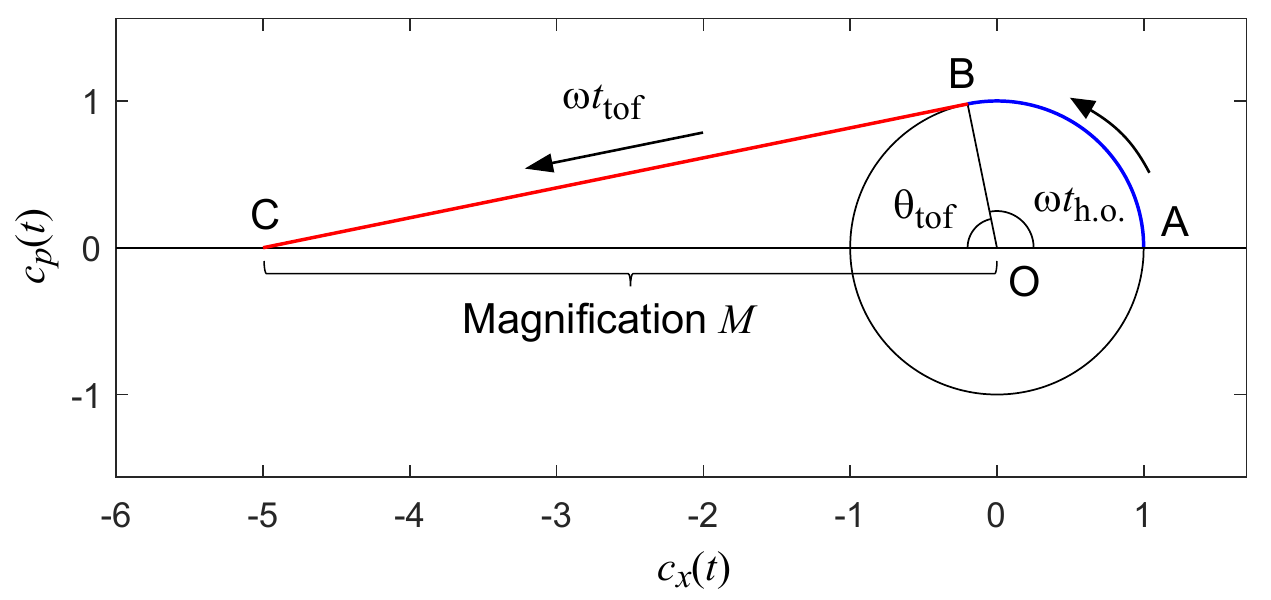}
		\caption{{\bf Graphical representation of the focusing condition.} Time evolution of the operator $\tilde{x}=c_x(t) \tilde{x}(0)+c_p(t)\tilde{p}(0)$ (with $c_x(t)$,$c_p(t)$  time-dependent coefficients),
		as a combination of the time-independent operators $\tilde{x}(0),\tilde{p}(0)$. The evolution starts at $A=(1,0)$ and performs a rotation in the HO for a duration $t_{\rm ho}$. Upon reaching point $B$, time of flight begins, as described by the straight line $BC$ (since during tof $\tilde{p}=\partial_t \tilde{x}={\rm const.}$) of length $\omega t_{\rm tof}$ (since at all times the velocity in this plane is $\omega$). To get $|c_x|=M$, $c_p=0$ one sees that $\omega t_{\rm ho}+\theta_{\rm tof}=\pi$ must hold, adding up to the half circle with $\theta_{\rm tof}=\arctan(\omega t_{\rm tof})$, and that $M=\sqrt{1+(\omega t_{\rm tof})^2}$ as the hypotenuse of the right-angled triangle 0BC. In the far field limit $\omega t_{\rm tof}\rightarrow \infty$, $\omega t_{\rm ho}\rightarrow\pi/2$ i.e. an exact $T/4$ pulse. %\note{KSe: hmm, das illustriert sie doch nicht wirklich, im Sinne, das es explizit gezeigt wird, sondern man muss es sich überlegen. Wie wäre: The far field limit (omega*ttof--> unendlich) would correspond to an excat T/4 pulse to position (cxt=0,cpt=1) followed by time of fligt - line parallel to cx-axs twowards - unendlich. --> ???}
		} \label{fig:focusing-condition}  
\end{figure}

We solve the Hamilton equations for the time-dependent $\tilde{x}=x\sqrt{\frac{m\omega}{\hbar}}$ and $\tilde{p}={\frac{p}{\sqrt{\hbar m\omega}}}$ operators (in the natural harmonic oscillator units of the harmonic trap with trapping frequency $\omega$ and mass $m$) in the Heisenberg representation:
\begin{equation*}
\partial_t 
\begin{pmatrix} 
\tilde{x} \\
\tilde{p} 
\end{pmatrix}
=\omega\begin{pmatrix} 
0 & 1 \\
-1 & 0 
\end{pmatrix}\begin{pmatrix} 
\tilde{x} \\
\tilde{p} 
\end{pmatrix}
\end{equation*}
This gives after a time $t_{\rm ho}$ in the HO:
\begin{gather*}
    \tilde{x}(t_{\rm ho})=\cos(\omega t_{\rm ho})\tilde{x}(0)+\sin(\omega t_{\rm ho})\tilde{p}(0)\\
\tilde{p}(t_{\rm ho})=\cos(\omega t_{\rm ho})\tilde{p}(0)-\sin(\omega t_{\rm ho})\tilde{x}(0).
\end{gather*}
After a time of flight expansion time $t_{\rm tof}$ it becomes:
\begin{equation*}
    \begin{aligned}
        \tilde{x}(t_{\rm ho}+t_{\rm tof}) & =\tilde{x}(t_{\rm ho})+\tilde{p}(t_{\rm ho})\omega t_{\rm tof}\\ & =\tilde{x}(0)\left[\cos(\omega t_{\rm ho})-\omega t_{\rm tof}\sin(\omega t_{\rm ho})\right]\\ & +\tilde{p}(0)\left[\sin(\omega t_{\rm ho})+\omega t_{\rm tof}\cos(\omega t_{\rm ho})\right]\\ & =\tilde{x}(0)M\left[\cos(\theta_{\rm tof})\cos(\omega t_{\rm ho})-\sin(\theta_{\rm tof})\sin(\omega t_{\rm ho})\right]\\ & +\tilde{p}(0)M\left[\cos(\theta_{\rm tof})\sin(\omega t_{\rm ho})+\sin(\theta_{\rm tof})\cos(\omega t_{\rm ho})\right]
    \end{aligned}
\end{equation*}
where we introduced the phase space rotation angle $\theta_{\rm tof}=\arctan(\omega t_{\rm tof})$ and $M=\frac{1}{|\cos( \theta_{\rm tof})|}= \sqrt{1+(\omega t_{\rm tof})^2}$ (Fig.~\ref{fig:focusing-condition}). Finally, one can write:
\begin{equation*}
\tilde{x}(t_{\rm ho}+t_{\rm tof})=M\left[ \tilde{x}(0)\cos(\omega t_{\rm ho}+\theta_{\rm tof}) + \tilde{p}(0)\sin(\omega t_{\rm ho}+\theta_{\rm tof})\right]
\label{eq:xoperator}
\end{equation*}

The focusing condition is given by 
\begin{equation}
   \tan(\omega t_{\rm ho})=-\omega t_{\rm tof} 
   \label{eq:focusing-condition}
\end{equation}
or equivalently by
\begin{equation*}
   \theta_{\rm tof}=\arctan(\omega t_{\rm tof})=-\omega t_{\rm ho}+n\pi,
\end{equation*} with $n$ integer.

This condition avoids a mixing of the initial momentum $\tilde{p}(0)$ into the final position $\tilde{x}(t_{\rm ho}+t_{\rm tof})$ and therefore reproduces the initial density distribution without distortion. The magnification of this imaging is given by
\begin{equation*}
    M=\sqrt{1+(\omega t_{\rm tof})^2}\approx \omega t_{\rm tof} 
    \label{eq:magnification*}
\end{equation*}
or equivalently by
\begin{equation*}
    M=\frac{1}{|\cos(\omega t_{\rm ho})|}. 
\end{equation*}
Note that we always state $M$ as a positive number although the imaging can be inverting. The focusing condition Eq.~(\ref{eq:focusing-condition}) is fulfilled when the evolution time in the harmonic oscillator. is close to an odd multiple of a quarter of the oscillation period $T=2\pi/\omega$. For values of $t_{\rm ho}\sim(1+4n)\frac{T}{4}$ values of $\cos(\omega t_{\rm ho}+\theta_{\rm tof})$ are negative and the image would then be inverted, while regular imaging is realized at $t_{\rm ho}\sim (3+4n)\frac{T}{4}$. In the far-field limit of the ToF expansion $\omega t_{\rm tof}\gg 1$, the magnification simplifies to $M\approx \omega t_{\rm tof}$ and the evolution time in the harmonic trap reduces to $t_{\rm ho}\approx T/4$. For typical magnifications $M>35$, $t_{\rm ho}$ deviates from $T/4$ by less than $2\%$.

Our large magnifications of up to $M=93$ are reached via the use of rather large trap frequencies. In the case that larger magnifications are desired, the scheme can easily be adapted. The magnification could be increased via longer time of flight expansion accessible via levitation or by adding an additional time evolution in an anti-confinement after evolution in the harmonic confining potential, in analogy to the proposal for magnification of the momentum distribution in ref.~\cite{Murthy2014S}. 

\subsection{Magnified imaging of the momentum distribution}

In contrast, when choosing $\cot(\omega t_{\rm ho})=\omega t_{\rm tof}$, one gets $\tilde{x}=M\tilde{p}_0$, i.e. one can measure the momentum distribution without distortion even for finite ToF expansion time $t_{\rm tof}$ but with a  magnification arbitrarily tunable in the range between $1$ and $\sim \omega t_{\rm tof}$, where $1$ is obtained with a pure $T/4$ evolution in the trap \cite{Tung2010S, Murthy2014S} and $\sim \omega t_{\rm tof}$ by a pure time of flight evolution. The advantage of avoiding the far-field approximation has to be balanced with possibly stronger interaction effects during the matter wave optics as discussed in Section~\ref{section:interactions}. 

\subsection{Invariance to time-dependent parameters}
We provide a proof that distortion free imaging is attainable also in the case where the trap confinement $\omega$ or the trap center are time dependent as long as the trap remains harmonic. This is relevant, e.g., for finite switching times of the magnetic trap providing the confinement, which is around $40\,$µs in our experiments. It is also relevant for magnification protocols during which the trap frequency is ramped up, which can additionally move the trap position due to gravitational sag. 

For a time dependent trapping frequency $\omega(t)$ and a time dependent center $c(t)$ we can write:
\begin{equation*}
\partial_t 
\begin{pmatrix} 
\tilde{x} \\
\tilde{p} 
\end{pmatrix}
=\begin{pmatrix} 
0 & \omega_0\\
-\omega^2(t)/\omega_0 & 0 
\end{pmatrix}\left[\begin{pmatrix} 
\tilde{x} \\
\tilde{p} 
\end{pmatrix} -\begin{pmatrix} 
c(t) \\
0 
\end{pmatrix} \right].
\end{equation*}
Where $\omega_0$ is the reference time-independent trap frequency entering the definition of the dimensionless operators.
We note that this is still valid in presence of an additional constant force like gravity, since it would just shift the trap center; it has the solution: 
\begin{equation}
\begin{pmatrix} 
\tilde{x}\\
\tilde{p}
\end{pmatrix} =U(t)\left[-\int_0^{t}U^{-1}(t')\begin{pmatrix} 
c(t') \\
0 
\end{pmatrix} dt'+\begin{pmatrix} 
\tilde{x}(0) \\
\tilde{p}(0)
\end{pmatrix} \right]
\label{eq:time_dep_ev}
\end{equation}
with $U(t)$ (right to left multiplication): 
\begin{equation*}
U(t)={\rm lim}_{dt\xrightarrow[]{}0}\Pi_{n=0}^{\frac{t}{dt}}\left[1+dt\begin{pmatrix} 
0 & \omega_0\\
-\omega^2 (n dt)/\omega_0 & 0 
\end{pmatrix}\right]
\end{equation*}

The first term of the sum in Eq.~(\ref{eq:time_dep_ev})  just shifts $\tilde{x}$ by a (time-dependent) real number (the image would be displaced). The evolution restricted to the space spanned by the linear combinations of $\tilde{x}(0)$ and $\tilde{p}(0)$ is then described just by
\begin{equation*}\partial_t 
\begin{pmatrix} 
\tilde{x} \\
\tilde{p} 
\end{pmatrix}
=\begin{pmatrix} 
0 & \omega_0\\
-\omega^2(t)/\omega_0 & 0 
\end{pmatrix}\begin{pmatrix} 
\tilde{x} \\
\tilde{p} 
\end{pmatrix},\end{equation*}
which is solved by 
\begin{equation*} \begin{pmatrix} 
\tilde{x} \\
\tilde{p}
\end{pmatrix}=U(t)\begin{pmatrix} 
\tilde{x}(0) \\
\tilde{p}(0)
\end{pmatrix}. \end{equation*}
The matrix $\begin{pmatrix} 
0 & \omega_0\\
-\omega^2(t)/\omega_0 & 0 
\end{pmatrix}$ implies a rotation with angular velocity $\partial_t \theta>\min(\omega^2(t)/\omega_0,\omega_0)$ which is always finite as long as $\omega(t)$ remains finite, and the additional time of flight can only rotate vectors $\theta_{\rm tof}<\frac{\pi}{2}$: from continuity it follows that it must exist an optimal time in the HO that ensures zero $\tilde{p}(0)$ component in $\tilde{x}$. In such cases, the expression for the magnification might be involved. The finite switching time of the magnetic trap of $40\,$µs is about $10\%$ of $T/4$ for the biggest $\omega/2\pi \approx 700\,$Hz, resulting in a deviation of similar magnitude of $M$ from the estimate $\omega t_{\rm tof}$. Therefore, we determine the magnification experimentally by comparing the measured lengths of lattice vectors with the expected ones.

\subsection{Derivation in the Schr\"odinger picture}
We provide for completeness the derivation for the magnification in the Schr\"odinger picture, which is completely equivalent to the Heisenberg's one but it is useful because it describes the evolution of the quantum field operators $\Psi^\dagger(r)$ illustrating that the matter wave optics also reproduces quantum correlations. This possibility will allow for future fundamental studies in single-atom resolved regimes with the quantum gas magnifier.

We make use of the generating function $G$ of the Hermite polynomials $h_n(x)$ given by
\begin{equation*}
    G(x,g)=e^{-\frac{1}{2}x^2+2xg-g^2}=\Sigma e^{-\frac{1}{2}x^2}h_n(x)\frac{g^n}{n!}.
\end{equation*}
Using the operator $\hat{O}_n=(\partial g)^n|_{g=0}$, one gets
\begin{equation*}
    \hat{O}_{n}G(x,n)=e^{-\frac{1}{2}x^2}h_n(x)=\psi_n(x)
\end{equation*}
with $\psi_n(x)$ being the $n^{th}$ eigenstate of the HO and with $x$ being the spatial coordinate in natural units. Up to a global phase, the time evolution in the HO $U{(t_{\rm ho})}$ can be described by 
\begin{equation*}
    U{(t_{\rm ho})}G(x,g)=G(x,ge^{-i\omega t_{\rm ho}})
\end{equation*}
which is proven by checking that $\psi_n$ picks up a phase\newline $\phi_{ho,n}=-n\omega t_{\rm ho}$ :
\begin{equation*} 
    U(t_{ho})\psi_n(x)=\hat{O_n}G(x,ge^{-i\omega t_{\rm ho}})=\psi_n(x)e^{-in\omega t_{\rm ho}}.
\end{equation*}

The time of flight evolution $U{(t_{\rm tof})}$ of $G$ can be described by 
\begin{equation*}
    U{(t_{\rm tof})}G(x,g)=F^{-1}(\sqrt{2\pi}\int dx'e^{-\frac{1}{2}x'^2+2x'g-g^2} e^{ikx'-ik^2\frac{T}{2}})
\end{equation*}
with $T=\omega t_{\rm tof}$ and $F^{-1}$ being the inverse Fourier Transform operator. It follows:
\begin{gather*}
    F^{-1}(\sqrt{2\pi}\int dx' e^{-\frac{1}{2}(x'-2g-ik)^2-\frac{1}{2}k^2-ik^2\frac{T}{2}+2ikg+g^2})\\
    =F^{-1}(e^{-\frac{1}{2}k^2+g^2-ik^2\frac{T}{2}+2ikg})\\
    =\sqrt{2\pi}\int dk  e^{-\frac{1}{2}(k^2 D-\frac{2ig}{\sqrt{D}}+\frac{ix}{\sqrt{D}})^2+g^2+\frac{1}{2D}(-4g^2-x^2+4xg)})
\end{gather*}
with $D=1+iT$. One gets, recalling $M=\sqrt{1+T^2}$: 
\begin{gather*}
    =\frac{1}{\sqrt{D}}e^{-\frac{x^2}{2D}+\frac{2xg}{D}+g^2(1-\frac{2}{D})}\\
    =\frac{1}{\sqrt{D}}e^{\frac{x^2}{2}\frac{iT}{1+T^2}}e^{-\frac{x^2}{2}\frac{1}{1+T^2}+\frac{2xg}{1+iT}-\frac{1-iT}{1+iT}g^2}\\
    =\frac{1}{\sqrt{M}}e^{-i\frac{\arctan(T)}{2}+\frac{iT}{2}(\frac{x}{M})^2}G(\frac{x}{M},g\cdot e^{-i\cdot \arctan(T)})
\end{gather*}
It follows:
\begin{gather*} 
    U(t_{\rm tof})\psi_n(x)=\hat{O}_nU(t_{\rm tof})G(x,g)=\\
    \frac{1}{\sqrt{M}}e^{-i\frac{\arctan(T)}{2}+\frac{iT}{2}(\frac{x}{M})^2}\psi_n(\frac{x}{M})e^{-in\cdot\arctan(T)}
\end{gather*}
One gets a normalization factor $\frac{1}{\sqrt{M}}$ and a $n$-independent phase $e^{i\phi(x,T)}=e^{-i\frac{\arctan(T)}{2}+\frac{iT}{2}{(\frac{x}{M})^2}}$. During time of flight, $\psi_n$ gets magnified by a factor $M$ and picks up a phase $\phi_{{\rm tof},n}=-n\cdot \arctan(T)$. The total time evolution of a generic wavefunction $\psi(x)=\Sigma_nc_n\psi_n(x)$ during the magnification protocol is then:
\begin{equation*}
    \\U(t_{\rm tof})U(t_{\rm ho})\psi(x)=\frac{1}{\sqrt{M}}e^{i\phi(x,T)}\Sigma_n c_n\psi_n(\frac{x}{M})e^{i\phi_n}
\end{equation*}
and the focusing condition Eq.~(\ref{eq:focusing-condition}) for the magnified imaging can then be obtained by requiring that
\begin{equation*}
    \phi_n=\phi_{{\rm ho},n}+\phi_{{\rm tof},n}=-n\cdot\left[\omega t_{\rm ho}+\arctan(\omega t_{\rm tof})\right]=-n\cdot \pi
\end{equation*}
As a consequence the even states ($\psi_n(x)$ with even $n$) don't pick up any phase and the odd states (with odd $n$) pick up a phase $\pi$, which is the reason for the inversion of the distributions. The phase factor $e^{\frac{iTx^2}{2M^2}}$ means that although density correlations $g_2(r)$ are reproduced via the scaling $g_2(r')=g_2(M\cdot r)$, phase correlations $g_1(r)$ have to be treated with care. This comes from the fact that the protocol presented here simply rescales the real space positions, but not the momentum operator. 

\section{Characterization of the quantum gas magnifier}

\subsection{Imaging resolution}

\begin{figure}%[ht]
		\includegraphics[width=0.95\linewidth]{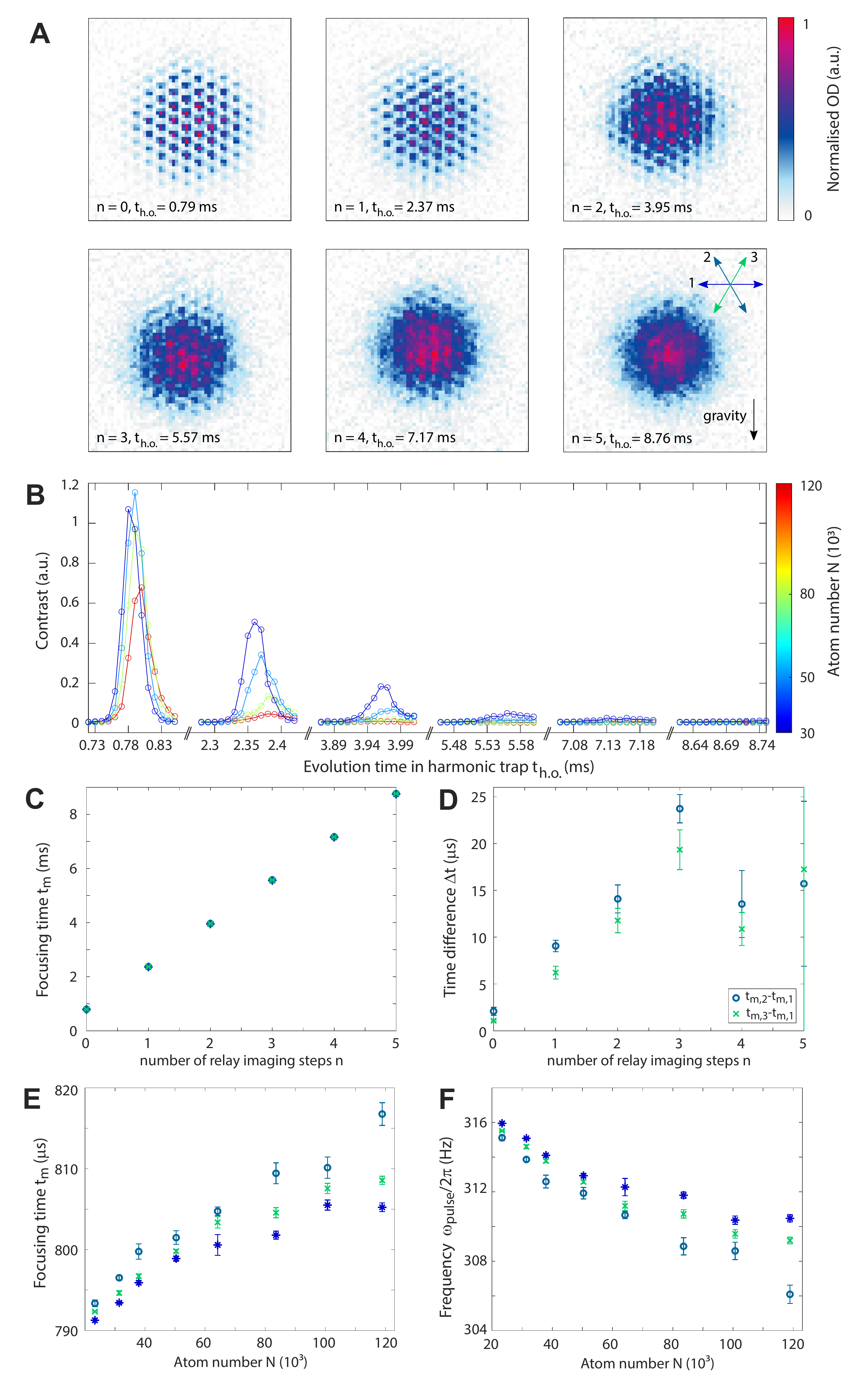}
        \caption{{\bf Image contrast of the quantum gas magnifier.} ({\bf A}) Images of the triangular lattice using increasing numbers of odd multiples in the harmonic trap: $(2n+1) T/4$ for $n=0,...,5$. The average atom number is 30,000 corresponding to the dark blue color in B. ({\bf B}) Lattice contrast versus the evolution time $t_{\rm ho}$. The colors specify the average atom number used in the images. ({\bf C}) Focusing times $t_m$ obtained as times of maximal contrast determined by fitting a Gaussian function to every contrast peak of the blue lines from B (blue stars: direction 1, teal circles: direction 2, green crosses: direction 3, for comparison of the directions see inset in A). ({\bf D}) Differences between the times of the two 1D lattice along direction 2 and 3 (not perpendicular to gravity) and the 1D lattice along direction 1 (perpendicular to gravity), illustrating the slight ellipticity along gravity ({\bf E}) Times of maximal contrast for $n=0$ and different atom numbers for the three 1D lattices. ({\bf F}) Corresponding effective trap frequencies, the decrease of which we attribute to mean field repulsion.}\label{fig:oddMultiples}	
	\end{figure}

In order to characterize the matterwave optics and the employed magnetic trap, we study the images for odd multiples of $T/4$, i.e. matterwave relay imaging by adding $T/2$ evolutions, in which all aberrations are amplified and can be better quantified. We define the contrast of the quantum gas magnifier as the integrated strength of the peaks around the reciprocal lattice vectors in the Fourier transformation of the measured densities normalized to the total atom number. Fig.~\ref{fig:oddMultiples}A shows how the lattice contrast of the images decreases for these multiples as expected. Fig.~\ref{fig:oddMultiples}B shows the contrast around the expected evolution times in the trap $t_{\rm ho}$ corresponding to $(2n+1) T/4$ for $n=0,...,5$. For a larger number of relay imaging steps $n$, the contrast cannot be recovered and the contrast also gets worse for larger atom numbers. This data is taken after removing the coherence between different tubes by ramping into a deep optical lattice. Without removing coherence, the contrast is lost much faster. 

From this data, we also obtain the precise time of maximal contrast, which we plot in Fig.~\ref{fig:oddMultiples}C. We find that the time of optimal contrast slightly depends on the direction in which the contrast is evaluated (Fig.~\ref{fig:oddMultiples}D) and we attribute this to the graviational sag, which slightly opens the trap along the vertical direction. The resulting ellipticity is only on the order of $1\%$ allowing us to focus the image in $x$ and $y$ direction simultaneously. The exact time of optimal contrast also slightly depends on the atom number as depicted in Fig.~\ref{fig:oddMultiples}E,F. We attribute this to a mean field repulsion, which effectively reduces the trap frequency of the external trap. 

\begin{figure}%[ht]
		\includegraphics[width=0.95\linewidth]{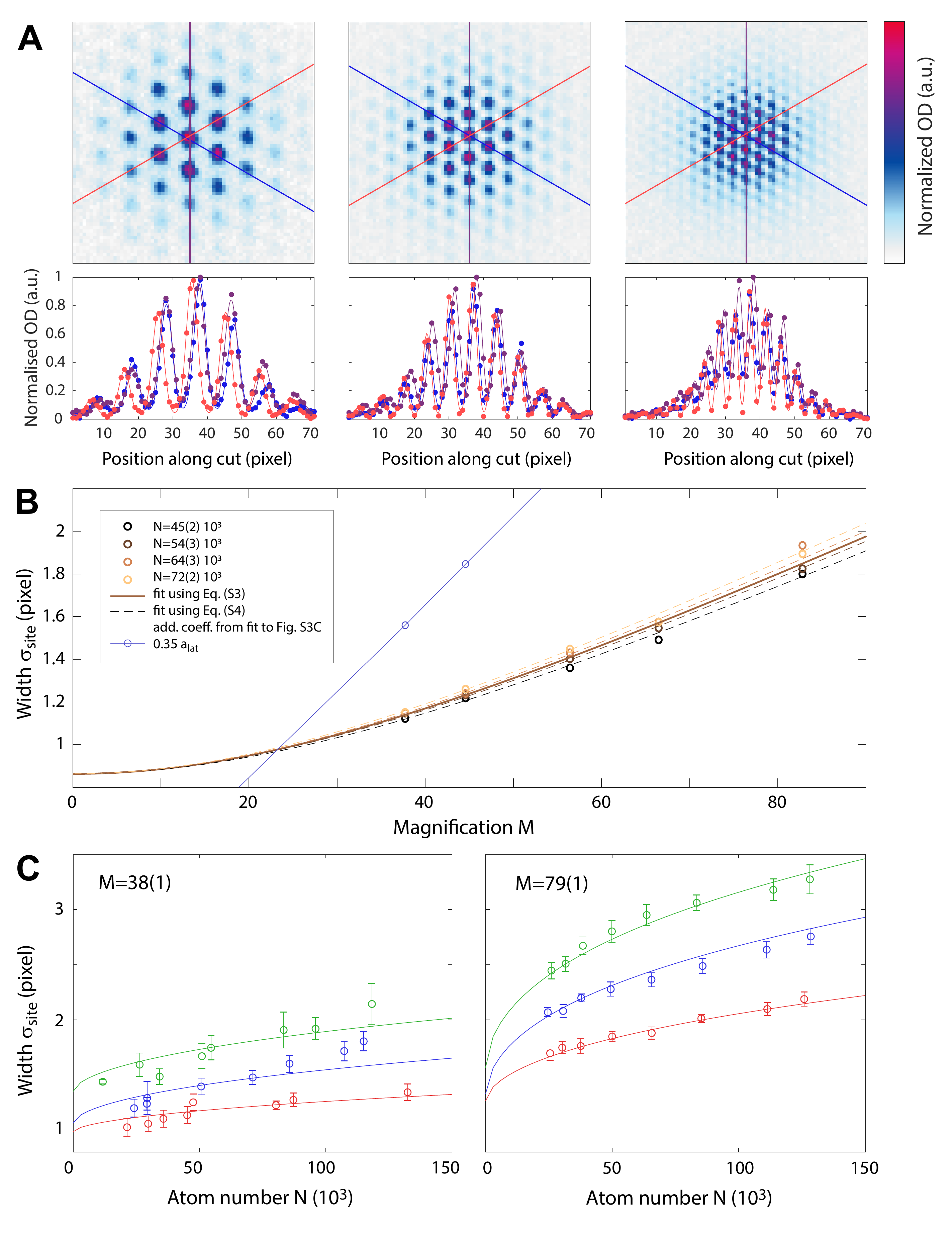}
        \caption{{\bf Resolution of the quantum gas magnifier.} ({\bf A}) Images with varying magnification (from left): $M=83(1)$, 56(1), 38(1). Beneath are cuts through the most populated rows along the three lattice directions, marked by correspondingly coloured lines in the images. The lines connecting the data points result from fitting a grid of Gaussian functions with a global $1/\sqrt{\rm e}$ radius $\sigma_{\rm site}$ for every lattice site. ({\bf B}) Fitted $\sigma_{\rm site}$ for five different magnifications and varying atom number. The brown line shows the fit according to Eq.~(\ref{eq:resolution}), independent of the atom number. The dashed lines describe the dependency from Eq.~(\ref{eq:fullResolution}), using the parameters $\sigma_{\rm lens}$ and $p_1$, $p_2$ and $p_3$ from C. The lattice sites can be resolved according to the Rayleigh criterion when $\sigma_{\rm site}<0.35 a_{\rm lat}$ (blue line). ({\bf C}) Gaussian width $\sigma_{\rm site}$ from fits to data along cuts in images as in Fig.~\ref{fig:oddMultiples}A for different atom numbers, the first three odd multiples of T/4 (red: n=0, blue: n=1, green: n=2) and two magnifications. The two line triplets are from a common fit of Eq.~(\ref{eq:fullResolution}) to the data in both panels in (C).}\label{fig:MOMresolution}
	\end{figure}

For a more quantitative measure of the resolution of the quantum gas magnifier, we fit a grid of Gaussians with global $1/\sqrt{\rm e}$ radius $\sigma_{\rm site}$ to a central cut of the images along the three lattice axes. We repeat this for several magnifications $M$, i.e. different trapping frequencies, and plot the resulting width versus magnification (Fig.~\ref{fig:MOMresolution}A,B). For magnifications above $M=35$, the lattice sites are well resolved according to the Rayleigh criterion with $\sigma_{\rm site}<0.35 a_{\rm lat}$. The experimental resolution $\sigma_{\rm site}$ can be reasonably well described as a convolution of the actual size of the wavefunction in the lattice site $\sigma_{\rm wf}$ multiplied by the magnification $M$ and the optical resolution of the absorption imaging $\sigma_{\rm opt}$: 
\begin{equation}
    \sigma_{\rm site}(M)=\sqrt{\sigma_{\rm opt}^2+(M \sigma_{\rm wf})^2}.
    \label{eq:resolution}
\end{equation} 
The brown line in Fig.~\ref{fig:MOMresolution}B shows this dependency, yielding $\sigma_{\rm opt}=5.2(2)\,\mu$m, $\sigma_{\rm wf}=118(3)\,$nm.

A more careful analysis should also include interaction effects during the matter wave optics. In order to increase these usually small effects, we repeat the analysis for different atom numbers $N$ and for odd multiples of $T/4$ (Fig.~\ref{fig:MOMresolution}C). A more complete description is then given by the heuristic model
\begin{equation}
\begin{aligned}
    \sigma_{\rm site}^2(M,N,n)= & \sigma_{\rm opt}^2+(M \sigma_{\rm wf})^2 +([2n+1]^{p_1} N^{p_2} M \sigma_{\rm int})^2 \\ & +([2n+1]^{p_3} \sigma_{\rm lens})^2
\end{aligned}
\label{eq:fullResolution}
\end{equation} 
with the parameters $\sigma_{\rm opt}=5.3(3)\,\mu$m, $\sigma_{\rm wf}=68(24)\,$nm, $\sigma_{\rm int}=4.2(4)\,$nm, $\sigma_{\rm lens}=0.42(46)\,\mu$m and the exponents $p_1=0.29(5)$, $p_2=0.33(5)$ and $p_3=1.6(7)$. Here $\sigma_{\rm int}$ takes into account the broadening due to interactions and $\sigma_{\rm lens}$ the single particle broadening. Note that the optical resolution $\sigma_{\rm opt}$ is given here as the $1/\sqrt{e}$-width of the point spread function, which corresponds to a Rayleigh resolution of about $r_0=(3.83/1.35) \sigma_{\rm opt}=15\,\mu$m. The dashed lines in Fig.~\ref{fig:MOMresolution}B show the fit of Eq.~(\ref{eq:fullResolution}) using $\sigma_{\rm lens}$, $p_1$, $p_2$ and $p_3$ from Fig.~\ref{fig:MOMresolution}C, yielding very similar results: $\sigma_{\rm opt}=5.2(1)\,\mu$m, $\sigma_{\rm wf}=75(10)\,$nm, $\sigma_{\rm int}=3.6(3)\,$nm. The optical resolution also agrees with the simpler fit of Eq.~(\ref{eq:resolution}), whereas $\sigma_{\rm wf}$ is overestimated in that case due to the negligence of interaction effects.

In contrast to the analysis above, the coherence was not removed for the nanoscale dynamics data of Fig.~4 and we therefore expect more interaction effects due to the residual coherence even in the relatively deep lattice. Indeed, we find about 17$\%$ of the atoms scattered into a constant background (compare the different color maps in Fig.~4C). These stronger interaction effects might also be related to the more complicated situation of a honeycomb lattice, in which scattering processes at the beginning of the ToF expansion can play a role for spin-mixtures \cite{Weinberg2016S, Thomas2016S}.

\subsection{Discussion of matter wave aberrations}

\begin{figure}%[ht]
		\includegraphics[width=0.75\linewidth]{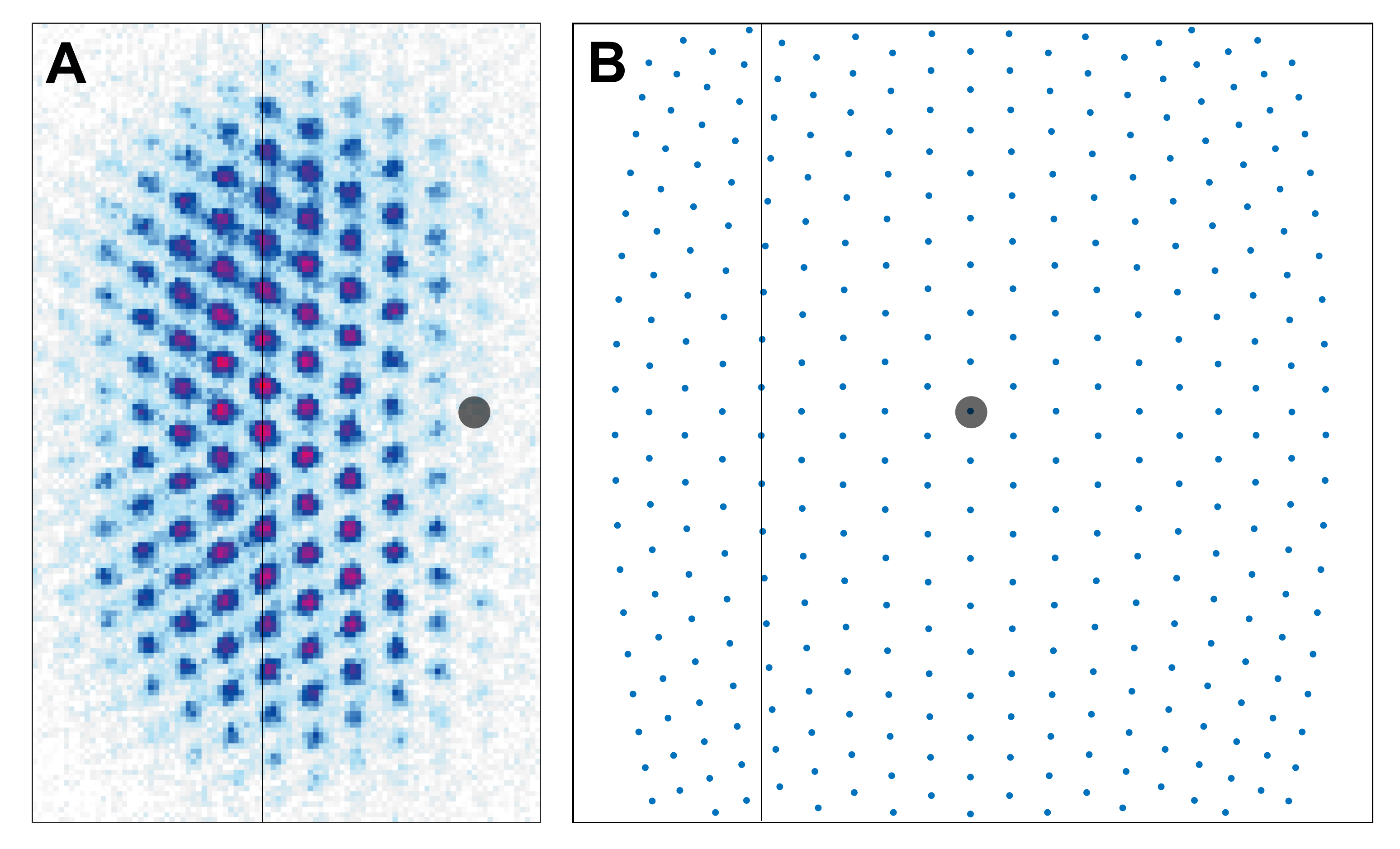}
        \caption{{\bf Aberrations due to trap anharmonicity.} ({\bf A}) Sum of two images with the cloud displaced in the different directions. For large trapping frequencies (here $\omega_{\rm ho}=2\pi \cdot 641\,$Hz) and atoms far from the center of the trap ('optical axis', grey circle), the anharmonicity of the trap gives rise to a distortion of the matter wave optics image, directly visible as a distortion of the imaged lattice . ({\bf B}) The distortion of the positions can be reconstructed by a simulation of the dynamics of a lattice of classical point particle.}\label{fig:anharmonic}	
	\end{figure}

The matter wave optics can also include single particle aberrations, e.g. due to anharmonicity of the matter wave lens.
The choice of a magnetic trap for the harmonic oscillator potential allows not only for a very smooth and isotropic trap, but also for smaller anharmonicities than in typical optical traps. The anharmonic (or aspheric) aberration can therefore be said to set a limitation on the usable field of view. For our magnetic trap, such aberrations become visible only for very large magnifications and for large systems or when displacing the cloud relative to the trap center. In order to control and characterize this anharmonicity we image a cloud shifted off-center in in a very strong magnetic trap with a trap frequency in the $x-y$ plane of $\omega_{\rm ho}=2\pi\cdot641\,$Hz, which results from a gradient $B_1=1.69 \cdot 10^4\,$G/m, a trap bottom $B_0=0.112\,$G and an anticurvature $B_2=7.1 \cdot 10^5\,$G/m$^2$ (Fig.~\ref{fig:anharmonic}). The small distortion of the image is no limitation when one is interested in the lattice site occupations. For an optical trap, it would be more difficult to avoid anharmonicity over a large field of view unless very large beam waists are used \cite{Kovachy2015S}.

The small effect of aberrations due to anharmonicity or other imperfections in magnetic traps is also supported by the observation in ref.~\cite{Hueck2018S} that the density distribution after a $T/2$ evolution in a harmonic trap is almost identical to the original distribution for the case of a non-interacting 2D Fermi gas. 

\subsection{Discussion of interaction effects} \label{section:interactions}

The matter wave optics imaging is initialized by projecting onto a non-interacting system. Quick reduction of the density is achieved here by switching off the optical lattice; other possibilities are a fast release of the transverse confinement in bulk 2D systems or a switching off of the interactions via a Feshbach resonance. For suppressing interaction effects during the matter wave optics, we typically remove the coherence by freezing in a deep lattice of $6\,E_{\rm rec}$ for 12~ms right before the magnification protocol. This avoids the density peaks, which otherwise arise during the $T/4$ evolution due to interference (Talbot revivals for short times and Bragg peaks at the end of the $T/4$ pulse). For the analysis of the density sector in this article, removing the coherence is unproblematic and interaction effects are then small.

	\begin{figure}%[ht]
		\includegraphics[width=0.8\linewidth]{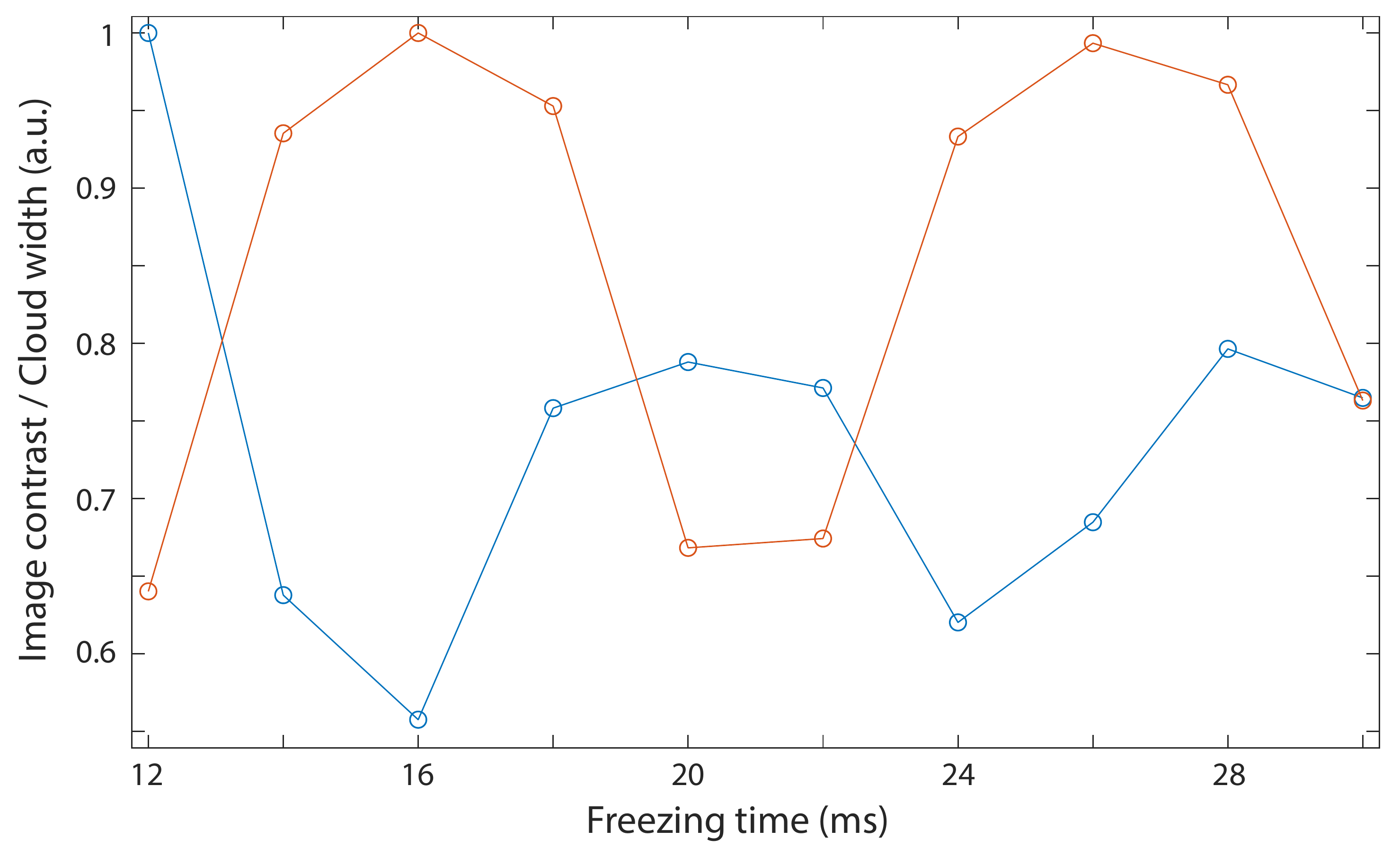}
        \caption{{\bf Contrast oscillation.} Width of the cloud along the line of sight of the usual imaging direction after the magnification protocol, as a function of the time after the excitation of a breathing mode in the direction perpendicular to the lattice (red curve).
        Broader clouds have a smaller widths during the first stages of the protocol and hence enhanced interactions. This leads to a smaller contrast of the quantum gas magnifier (blue curve).}\label{fig:contrastOscillation}
	\end{figure}

When ramping up the lattice intensity for the coherence removal procedure one can excite breathing oscillations along the transverse direction, over which the signal is integrated. This can increase the density during the $T/4$ evolution and therefore reduce the contrast due to interaction effects. The resulting oscillation of the contrast (Fig.~\ref{fig:contrastOscillation}) fits well to a breathing mode for the independent characterization of the transverse trapping frequency of the optical lattice. This analysis shows that the dynamics along the transverse direction is very important for the size of interaction effects. Therefore a protocol, which leads to a rapid expansion in the transverse direction without bringing the cloud out of the depth of focus of the optical imaging system, can be beneficial \cite{Murthy2014S}. 

Our characterizations provide a benchmark for the capability of the quantum gas magnifier concept and are also relevant for matter wave optics for imaging momentum space. This analysis demonstrates the full control of the quantum gas magnifier even including interaction effects for Rb-atoms and typical magnifications $M=50-90$. For other elements like $^{7}$Li or $^{39}$K the use of appropriate Feshbach resonances will even allow further extensions of the magnifier as interaction effects can be fully switched off prior to magnification and imaging. Vice versa interaction effects can be studied with the system with careful prior calibration.

Alternatively, interaction effects during the magnification protocol would be sufficiently suppressed for the small background scattering lengths at low magnetic fields, allowing to combine the use of Feshbach resonances for the initial system of interest with a tight magnetic trap for the matter wave optics.

In addition, we note that while possible interaction effects during the matter wave optics protocol have to be treated with care, the magnification and its resulting low densities completely remove interaction effects during the optical imaging, such as saturation effects at high optical densities \cite{Chomaz2012S} or light-assisted collisions in tight traps. This avoids distortions of the obtained density profile and reduces systematic effects on the absolute atom number calibration of absorption imaging. 

%%%%%%%%%%%%%%%%%%%%%%%%%%%%%%%%%%%%%%%%%%%%%%%%%%%%%%%%%%%%%%
%%%%%%%%%%%%%%%%%%%%%%%%%%%%%%%%%%%%%%%%%%%%%%%%%%%%%%%%%%%%%%

\section{System preparation and data evaluation}

\subsection{Optical lattice setup} 

Our optical lattice setup consists of three running waves of wave vector ${\bf k}_i$ with $\lvert {\bf k}_i \rvert=2\pi/\lambda$ intersecting under an angle of $120^\circ$. Depending on the polarization of the beams we obtain either a triangular lattice (linear polarization perpendicular to the lattice plane), a honeycomb lattice (linear polarization in plane) \cite{Becker2010S} or a boron-nitride lattice (suitable elliptical polarization of the lattice beams \cite{Flaschner2016S} as in this work or using spin-dependent light shifts \cite{Soltan-Panahi2011S}). 

The resulting potential can be written as
\begin{align}
	V_{\rm lat 2D}({\bf r}) =& \sum_{i>j} \sqrt{V_{\rm lat}^{(i)} V_{\rm lat}^{(j)}} \\
    \times [ &\cos^2(\theta) \cos\left(\left({\bf k}_i - {\bf k}_j\right){\bf r} + \alpha_i - \alpha_j \right) \nonumber\\
	- &2 \sin^2(\theta) \cos\left(\left({\bf k}_i - {\bf k}_j\right){\bf r} \right) ] \nonumber
\end{align}
where the $V_{\rm lat}^{(i)}$ are proportional to the intensities of the lattice beams. $\theta$ is the angle of the polarization (long half axis) with respect to the lattice plane, $\alpha_i$ is the relative phase between the s and p components of the polarization for beam $i$. We neglected the phases of the beams with respect to each other because they only result in a global shift of the lattice. If we just name a single lattice depth, then all $V_{\rm lat}^{(i)}$ are equal. The boron-nitride lattice in Fig.~3 uses $\theta = 9^{\circ}$ and ${\bf \alpha} = (0, 120^{\circ}, 240^{\circ})$ yielding an energy offset between the A and B sublattice quantified by the tight-binding parameter $\Delta_{AB}$ \cite{Flaschner2016S}. Note that the triangular lattice has a much larger barrier between nearest neighbours than the honeycomb or boron-nitride lattice for the same laser intensities~\cite{Windpassinger2013S}. 

\subsection{Read-out of lattice site populations} \label{ssec:pop-comp}

\begin{figure}
	\includegraphics[width=0.65\linewidth]{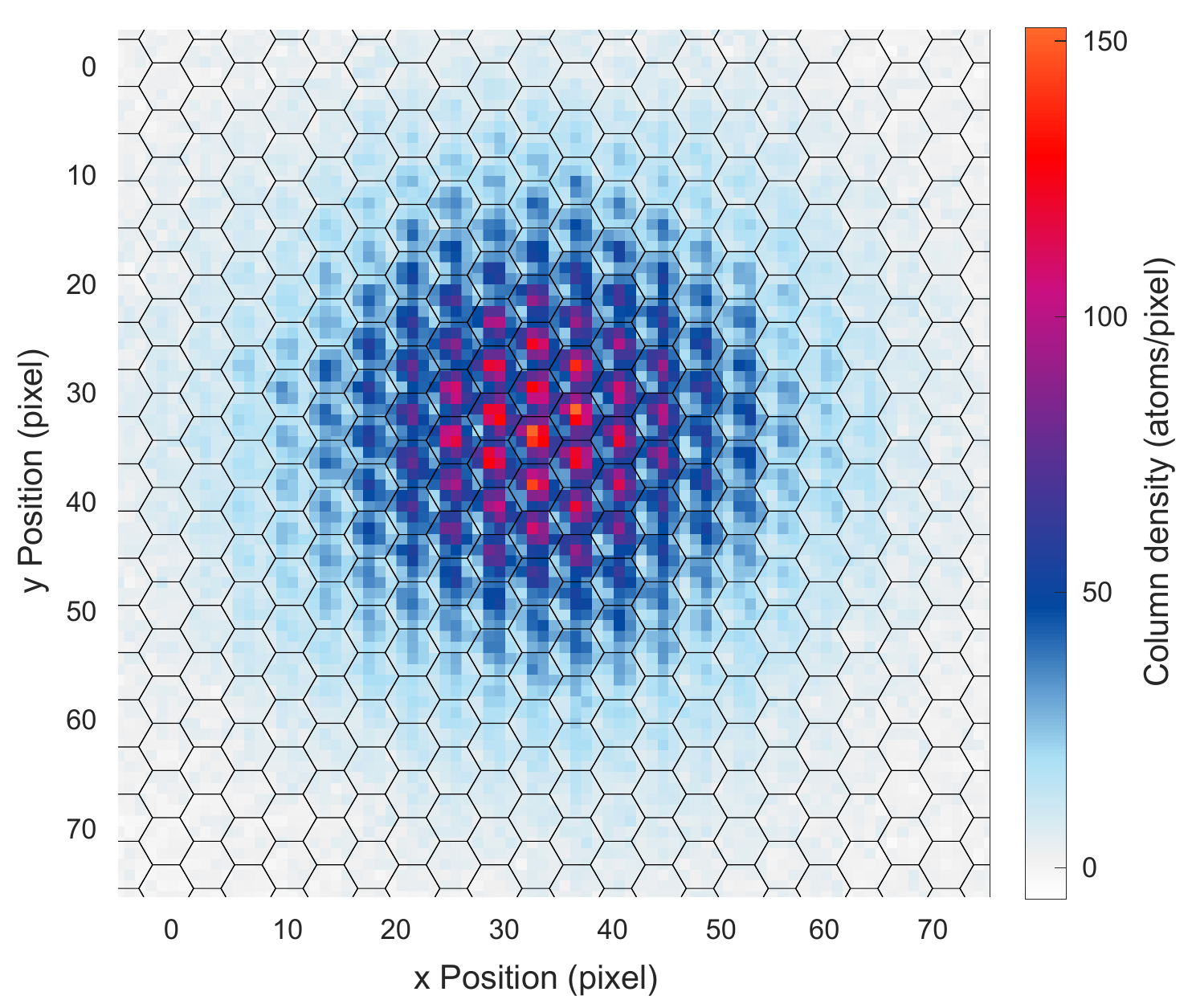}
        \caption{{\bf On-site population computation.} The raw density distribution underlying Fig.~2B is shown. The Wigner-Seitz masks used to determine the on-site populations are plotted as solid lines.}\label{fig:masks}	
\end{figure}

For several experiments only the total population of the lattice sites is of interest. We extract these by first fitting a triangular lattice to the data and subsequently summing up the signal in the Wigner-Seitz cells around the individual sites as explained in the following. The lattice constant $a_{\rm lat}$ in pixels is determined by integrating the density of individual images along a real space lattice vector yielding a 1D profile with lattice constant $a_{\rm 1D}$, which is obtained from a fit with the heuristic function $A\exp (-(x-x_0)^2/(2\sigma^2)) (\cos(\pi x/a_{\rm 1D} + \phi)^2 + \Delta)$. Finally, the lattice constant is deduced from the average fit parameter from two different such directions as $a_{\rm lat} = 2a_{\rm 1D}/\sqrt{3}$. Next, the spatial phase of the lattice is determined by multiplying the image with a mask that removes the signal from pixels at a certain radius around the sites of a triangular lattice with the lattice constant determined beforehand. The phase of this mask is varied and the configuration minimizing the remaining density is considered the lattice phase. The final step is to determine the population of each lattice site by summing over the Wigner-Seitz cell around the lattice site. To minimize discretization errors the pixels of the camera are subdivided such that the radius of the cell is about 10 subpixels. For an example image with non-discretized Wigner-Seitz masks see Fig.~\ref{fig:masks}. 

For the lattices with two-atomic basis we slightly adjust the algorithm for lattice phase determination by maximizing the density which is not masked thus locating the centers of the honeycombs. 

\subsection{Lattice phase drifts}

For our hexagonal lattice setup composed of three laser beams in two dimensions, phase shifts of the lattice beams  only lead to a translation of the whole lattice potential, but not to a change of the lattice geometry \cite{Petsas1994S}. We verify that such phase drifts are not a problem on the time scale of the experiments presented here by measuring the position drift of the atomic cloud's center of mass in a very deep optical lattice. We find that the cloud position moves and scatters by less than one lattice site peak-to-peak within $6\,$s hold time. We checked in a previous set of measurements where we deliberately move the lattice, that the lattice is deep enough to be able to drag the atoms along. Shot-to-shot lattice drifts exceed one lattice site (cycle time of 30~s).

Our characterization of the slow phase drifts is compatible with recent direct measurements of triangular lattices using quantum gas microscopes \cite{Yamamoto2020S, Yang2021S}. The drifts can be further reduced to one lattice site per minute in a setup with a single, refolded lattice beam \cite{Yang2021S}. In our case, the three beams go through separate optical fibers, a setup in which phase locks have been implemented to stabilize the phase \cite{Becker2010S}. 
From our characterization, we conclude that a phase lock is not necessary. The random lattice phase between individual images can be easily taken into account by identifying the phase. For data evaluation in the main text, we determine the lattice position for every experimental image via a fit routine as described in section~\ref{ssec:pop-comp}. Note that the envelope of the atomic density is given by the position of the magnetic trap and is therefore not affected by lattice phase drifts. 

\subsection{Bimodal fits of density profiles}

The lattice-gas profiles can be described by a bimodal model. Since we are considering the on-site populations only, the presence of the lattice can be included by a renormalization of the interaction constant \cite{Pitaevskii2016S} $g_{\rm eff} = g \times A_{\textrm{WS}} / (2\pi \sigma^2)$ and otherwise using a continuum formalism. Here, $A_{\textrm{WS}}$ is the area of the Wigner-Seitz cell, $\sigma$ the on-site radial oscillator length and $g = 4\pi \hbar^2 a_{\rm sc}/m$ the interaction constant, computed from the scattering length $a_{\rm sc} \approx 100$ Bohr radii and the mass $m = 87$u. The on-site radial oscillator length is computed as $\sigma = \sqrt{\hbar/(m\omega_{\rm onsite})}$ from the lattice depth using $\hbar\omega_{\rm onsite} = 3\sqrt{2 V_{\rm lat} / E_{\rm rec}} E_{\rm rec}$. The data in Fig.~2 is taken with a lattice depth of $V_{\rm lat} = 1 E_{\rm rec}$.

The condensed atoms are described by a 3D Thomas-Fermi profile integrated along line of sight, 
\begin{equation}
    n_{\rm BEC}(x,y) = \int{\rm d}z ~ \frac{15}{8\pi} \frac{N_{\rm BEC}}{R_\rho^2 R_z} \left(1 - \frac{\rho(x,y)^2}{R_\rho^2} - \frac{z^2}{R_z^2}\right) \rm{.}
\end{equation}
The fit parameters here are the center of the cloud $x_0$, $y_0$ resulting in $\rho(x,y)^2 = (x-x_0)^2 + (y-y_0)^2$, the in-plane Thomas-Fermi radius $R_\rho$ from which the out-of-plane radius $R_z$ is deduced via a computed aspect ratio, and the number of atoms in the BEC $N_{\rm BEC}$. In fact, only for the lowest evaporation frequency, where the BEC is very distinct from the thermal part, $N_{\rm BEC}$ and $R_\rho$ are fitted independently. For all other fits we compute the Thomas-Fermi radius from the number of condensed atoms using the expected scaling $R_\rho = \gamma N_{\rm BEC}^{1/5}$ with $\gamma$ determined as its mean value from the fits at lowest evaporation frequency. We obtain $\gamma = 0.354\,$µm which agrees excellently with the expected value $\gamma_{\rm theo}=0.352\,$µm obtained from~\cite{Pethick2002S}
\begin{equation}
    \gamma_{\rm theo} = 15^{1/5}\sqrt{\frac{\hbar\bar{\omega}}{m\omega_{system}^2}}\left(\frac{g_{\rm eff}}{g}\frac{a_{\rm sc}}{\bar{a}}\right)^{1/5}.
\end{equation} 
Here $\omega_{\rm system} = 2\pi \cdot 305\,$Hz, $\bar{\omega} = (\omega_{\rm system}^2 \omega_z)^{1/3}$, $\omega_z = 2\pi \cdot 29\,$Hz and $\bar{a} = \sqrt{\hbar/(m\bar{\omega})}$.

\begin{figure}%[ht]
		\includegraphics[width=0.95\linewidth]{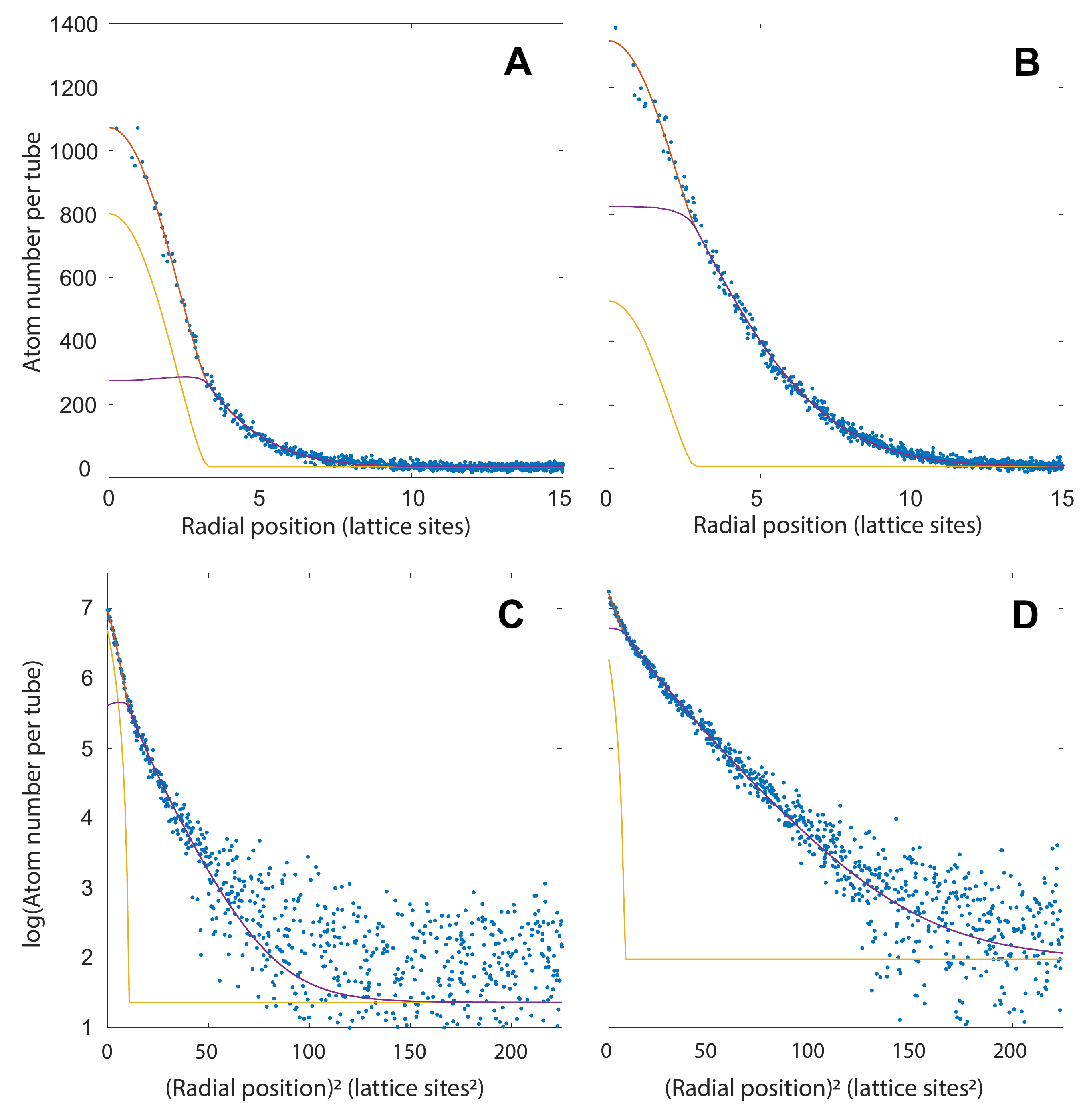}
        \caption{{\bf Radial density profiles} ({\bf A} and {\bf B}) Radial profiles of partially-condensed BECs in a lattice as shown in Fig.~2C,D. The examples are prepared by end points of the radio frequency evaporation of 85~kHz (A) and 105~kHz (B) and a hold time at the final evaporation frequency of 1~ms. ({\bf C} and {\bf D}) Same data as in (A) and (B) plotted with natural logarithmic $y$-axis and quadratic $x$-axis resulting in the Gaussian wings appearing as a straight line.}\label{fig:radialDensityProfiles}	
	\end{figure}

The thermal density distribution is described in a semi-ideal approach, i.e. as an ideal gas in a potential $V(x) = V_{\rm trap}(x) + V_{\rm BEC}(x)$ given by the external trap $V_{\rm trap}(x)$ and the repulsion from the condensed atoms $V_{\rm BEC}(x) = 2 g_{\rm eff} n_{\rm BEC}(x)$. In semi-classical approximation the ideal Bose gas density distribution is given by~\cite{Pethick2002S}
\begin{equation}
    n_{\rm th}(x) = g_{3/2}(\exp(-\beta(V(x) - \mu))) / \lambda_T^3    
\end{equation} 
with $g_{n}(x) = \sum_{i > 0} x^i/i^n$ and $\lambda_T = \hbar \sqrt{2\pi / (m k_B T)}$. Additionally we allow for a small offset which we substract when determining atom numbers. The fit is performed on the 2D density distribution and both the data and the fit function are subsequently plotted as radial profiles. Fig.~\ref{fig:radialDensityProfiles} shows the radial density profiles from Fig.~2 of the manuscript along with a plot of the logarithm of the density versus the square of the radius, which yields a straight line in the thermal wings. This plot shows the excellent agreement between data and fit and also makes the change of the slope at the onset of the BEC fraction more visible.

\subsection{Interaction shift and finite size shift}

Interactions are known to shift the critical temperature for the BEC transition with a sign depending on the trapping geometry. For a 3D harmonic trap in mean field approximation the shift is negative and given by~\cite{Giorgini1996S, Pethick2002S} 
\begin{equation}
    \Delta T_c/T_c\approx -1.33 \frac{g_{\rm eff}}{g} \frac{a_{\rm sc}}{\bar{a}} N^{1/6}
\end{equation} 
predicting a shift of about $-0.24$ for the typical atom number of the condensed samples of $N = 5 \cdot 10^4$, which is larger than the measured shift of $-0.099(4)$. However, for interactions of this strength the mean-field approximation overestimates the shift~\cite{Smith2011S}. Note that we are not aware of a prediction for our 2D-3D crossover geometry of an array of tubes. Our measurements thus set a benchmark for future theoretical studies on the interesting setting of Josephson junction arrays.

We also recall the prediction for the finite size shift of the critical temperature for a 3D harmonic trap. For an isotropic harmonic trap with trap frequencies $\omega_x$, $\omega_y$, $\omega_z$ and their geometric mean $\bar{\omega}=(\omega_x \omega_y \omega_z)^{1/3}$ and arithmetic mean $\omega_{\rm m}=(\omega_x + \omega_y + \omega_z)/3$, the shift is given by~\cite{Giorgini1996S, Pethick2002S}.
\begin{equation}
    \Delta T_c/T_c\approx -0.73 \frac{\omega_{\rm m}}{\bar{\omega}}N^{-1/3}.
\end{equation}  
With our trapping frequencies of $2\pi\cdot\,$(305, 305, 29)\,Hz, the anisotropy factor is $\omega_{\rm m}/\bar{\omega}=1.53$ and the expected shift is $-0.03$ for our atom number of $N\approx 5 \cdot 10^4$, i.e. much smaller than observed. Note that both interactions and finite size effects can contribute to the shift.

The observed smoothing over a range of almost 0.2 in rescaled temperature is only expected for much smaller atom numbers in the case of a 3D harmonic trap \cite{Ketterle1996S}. We therefore conclude that finite size effects are strongly enhanced in our 2D-3D crossover geometry of an array of tubes.
We have verified that the small condensate fractions involved in the smoothened transition do not arise from fit artifacts of the bimodal profile to the density profiles. The good agreement with the curve for the visibility shown in Section~\ref{section:visibiltiy} is further evidence that the signal is physical and demands for further theoretical studies.

\subsection{Theoretical description of the density of states}

\begin{figure}%[ht]
		\includegraphics[width=0.95\linewidth]{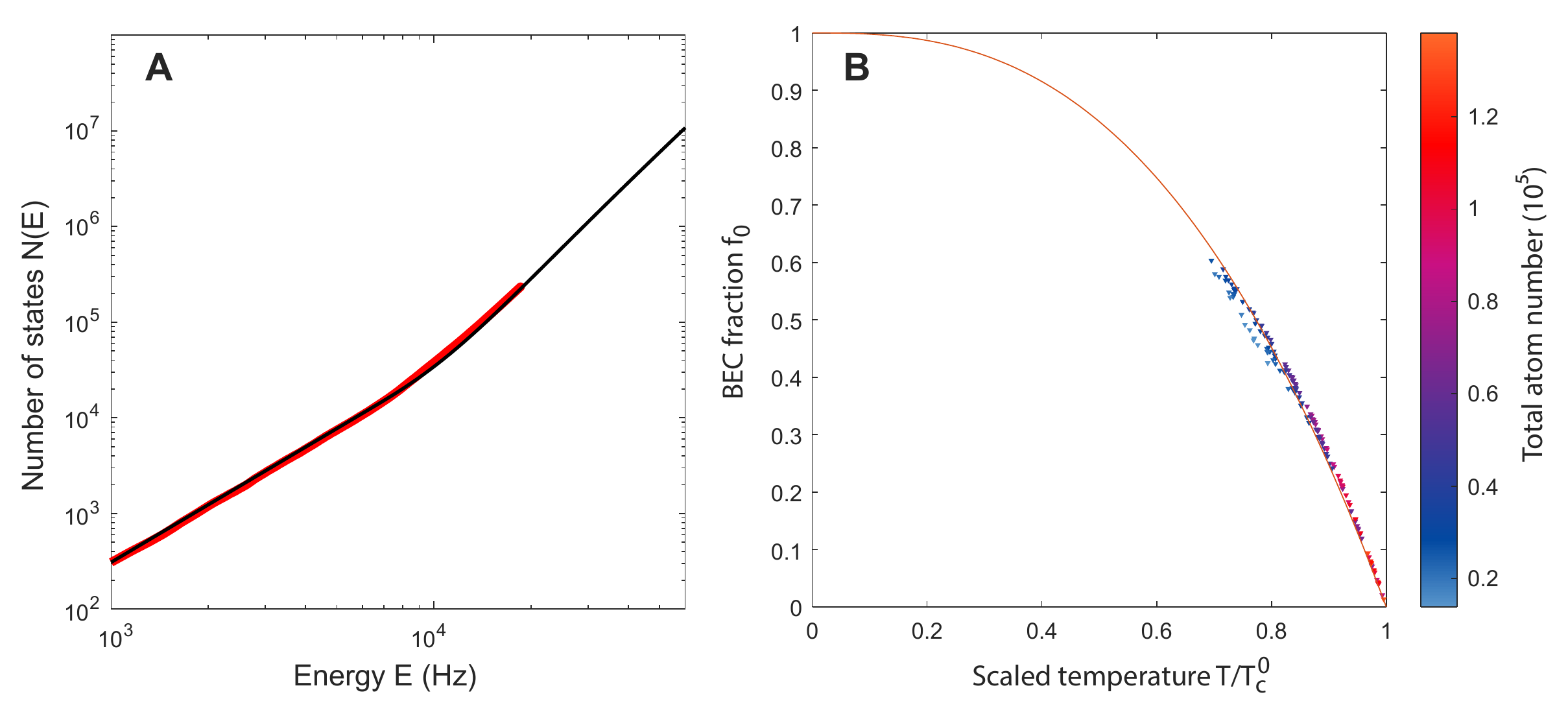}
        \caption{{\bf  Numerical evaluation of the density of states}. ({\bf A}) Number of states $N(E)$ up to energy $E$ for a triangular lattice of tubes with tight external confinement: the analytical approximation of Eq.~(\ref{eq:DOS-analytical}) (black curve) reproduces the numerical diagonalization of the system (red points). The crossover between $\alpha=2$ and $\alpha=3$ in the relevant energy range can be seen in the slope change in the log-log plot. ({\bf B}) Theoretical values of $f_0$ for the experimentally accessed parameters computed using the analytical approximation. The orange curve is a fit of the form $f_0 = 1 - (T/T_{\rm c}^0)^\alpha$ as explained in the main text.}\label{fig:densityOfStates}	
	\end{figure}

We compare our data of the thermal-to-BEC phase transition to non-interacting calculations based on the density of states. To this end we compute the Hamiltonian matrix for our trap in position basis and diagonalize it. In the numerical spectrum we clearly recognize a crossover between two power laws as a slope change in the log-log plot of Fig.~\ref{fig:densityOfStates}A. The asymptotes of this crossover can be understood using analytical considerations. 

The high energy limit coincides with the the well-known spectrum of a 3D harmonic trap resulting in 
\begin{equation}\label{eq:DOSho}
    N(E) = \frac{1}{6}\left(\frac{E}{\hbar \bar{\omega}}\right)^3
\end{equation} 
states up to energy $E$. This is due to the fact that the gaps between higher bands are negligible compared to the band widths. So we have to count separately the first band states and harmonic oscillator-like states.

For energies $E<\Delta_g$, where $\Delta_g$ is the first band gap, only states of the first kind are relevant. Here, the tunnel coupling $J = h \cdot 12$~Hz is negligible compared to the offset introduced by the external trap which is $\Delta = 1/2 m \omega_{\rm syst}^2 a_{\rm lat}^2 = h \cdot 200$~Hz for a site in the centre compared to a nearest neighbour. Hence the spectrum is given by
\begin{equation}
    E_{ijk} = 1/2 m \omega_{\rm syst}^2 r_{ij}^2 + (k+1/2)\hbar\omega_z    
\end{equation}
with $r_{ij}$ being the distance of the lattice site indexed $ij$ from the trap centre and $k$ is the index for the $z$ direction. A lengthy calculation leads to $N(E) = (E/E_0)^2$ with $E_0 = \sqrt{\hbar A_{\rm WS} m \bar{\omega}^3/\pi} = h \cdot 57$~Hz.

We can therefore find an approximation of the numerical result by the Ansatz 
\begin{equation}
    \label{eq:DOS-analytical}
    N(E)=\left(\frac{E}{E_0}\right)^2+{\rm max}\left(\frac{1}{6}\left(\frac{E-\hbar\Delta_g}{\hbar \bar{\omega}}\right)^3,0\right)  
\end{equation} 
where $\Delta_{\rm g}$ is obtained from a simulation without external trap. This analytical model fits very well to the exact diagonalization up to the numerically accessible energies (Fig.~\ref{fig:densityOfStates}A) while asymptotically reaching the known analytic limit of Eq.~(\ref{eq:DOSho}) for high energies. 

From $N(E)$ we obtain the density of states $g(E) = {\rm d}N/{\rm d}E$ which in turn allows to numerically compute the critical temperature $T_{\rm c}^0(N)$~\cite{Pethick2002S}. We find that for the experimentally relevant parameters the condensate fraction can be approximated by  $f_0 = 1 - (T/T_{\rm c}^0)^\alpha$ as obtained by assuming a density of states $g(E) = C_\alpha E^{\alpha - 1}$~\cite{Pethick2002S}. To determine the exponent for the range of energies probed by the experiment we compute the non-interacting prediction for our data points from Fig.\,2E in the main text and fit it with the mentioned functional form resulting in $\alpha = 2.69(1)$ (Fig.~\ref{fig:densityOfStates}B).

\subsection{Comparison to time-of-flight data}\label{section:visibiltiy}

\begin{figure}%[ht]
		\includegraphics[width=0.95\linewidth]{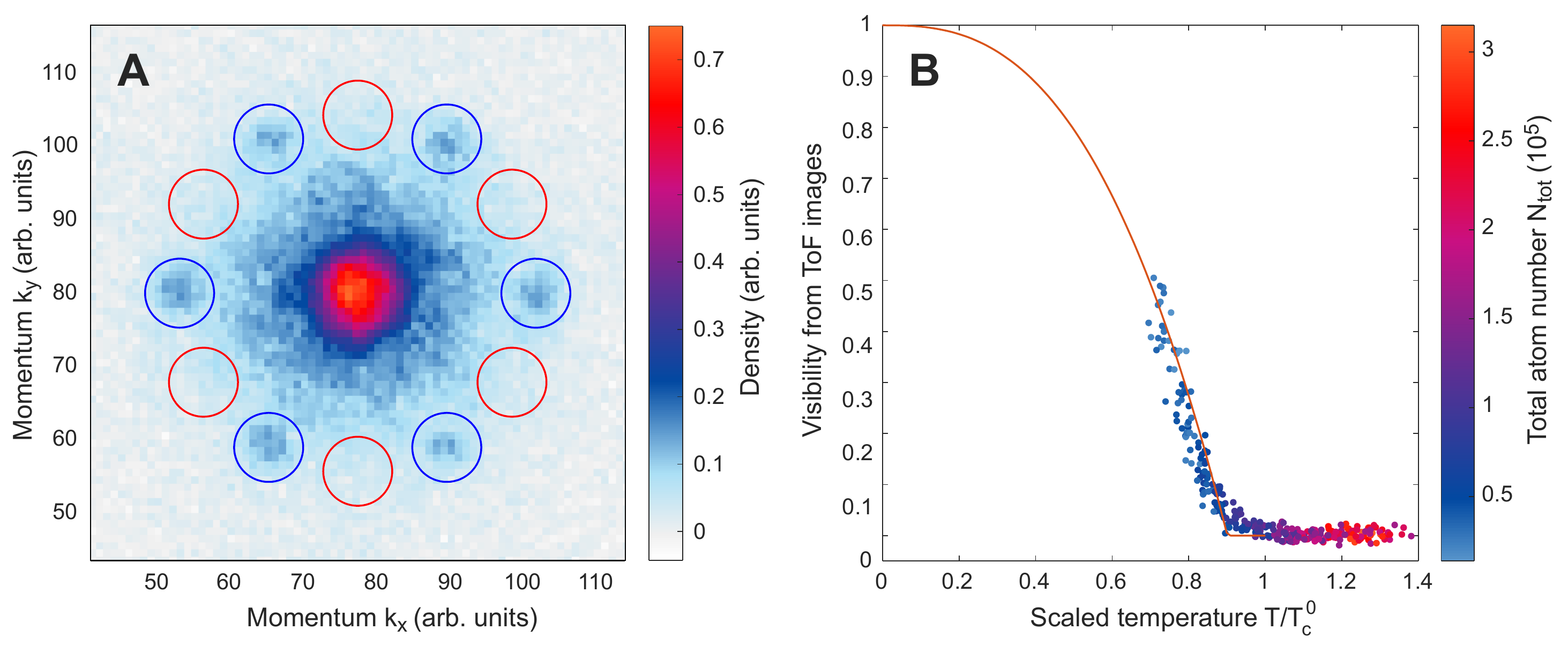}
        \caption{{\bf Visibility across the BEC phase transition.} ({\bf A}) Density after time of flight for the parameters of Fig.~2A in the main text showing the momentum space density with Bragg peaks reflecting the coherence in the system. The visibility is computed as the difference of the density in the blue circles and red circles compared to their sum. ({\bf B}) Visibility of the time-of-flight images of the lattice gas as a function of the reduced temperature $T/T^0_{\rm c}$ where $T^0_{\rm c}$ is the scaling temperature computed from the corresponding real space images in Fig.~2. The line is the fit to the experimental condensate fractions $f_0$ from the main text for reference.}\label{fig:visibility}	
	\end{figure}

For comparison, we also take momentum space images from ToF expansion at the same parameters and evaluate their visibility \cite{Gerbier2008S}, which is a measure of coherence in the system. We use circular masks around the Bragg peaks, the radius of which is identical to the Thomas-Fermi radius of a  fit (Fig.~\ref{fig:visibility}A). We plot the visibility as a function of $T/T^0_{\rm c}$ as obtained from the corresponding real space data (Fig.~\ref{fig:visibility}B). We plot the theory curve for the condensate fraction as a guide to the eye. This comparison shows that the real space and momentum space images give a compatible description of the system.

The visibility and the condensate fraction vanish for the same temperatures (see Figs.~2F and~\ref{fig:visibility}), indicating that no states with spatial coherence other than the ground state have significant population. This is due to the fact that the temperatures are high compared to the level spacings. This is in contrast to 3D optical lattices around unit filling, where the critical temperatures are much smaller, resulting in a finite visibility also for the case of vanishing condensate fraction~\cite{Trotzky2010S, Cayla2018S}.

\subsection{Details on magnetic resonance addressing}

In order to engineer the density distributions shown in Fig.~3, we used a constant trap frequency $\omega_{\rm addressing}/2\pi=543\,$Hz and different trap shifts and RF sequences. By shifting the magnetic trap perpendicularly to a real space lattice vector by around 14\,µm, corresponding to approximately twice the system diameter, the curvature of the equipotential lines becomes negligible and the density patterns created by addressing exhibit straight edges. In Fig.~3 the trap center resonance frequency is $\omega_c = 2\pi \cdot 108\,$kHz for all images. The trap is shifted by 14.1\,µm for the first and third image, by $15.7\,$µm for the second image and not shifted for the fourth to sixth image, but always shifted back to the position of the atoms before imaging. For the third image a constant RF pulse of 360\,kHz is turned on for 200\,ms. For the first image, an RF ramp from 360 to 290\,kHz is used, leading to the depletion of all lattice sites from the center of the cloud towards the center of the shifted magnetic trap. Here, for the same RF ramp time ($200\,$ms) we ramp over a wider range and therefore have to compensate the reduced time by which the resonance condition is met at each position by increasing the RF amplitude.In all protocols, Fourier broadening is negligible. Lattice phase fluctuations from shot to shot lead to one or two partially-depleted rows in most images. The second image in Fig.~3 is created by applying two RF ramps. In this case the trap was shifted further to the side resulting in a higher energy difference to the target $F=2$, $m_{F}=1$ state and thus we used ramps from 420 to 486 kHz and from 494 to 540 kHz with 200 ms each to target all sites except for the center line. For the fourth to sixth image 100 ms were used as the RF duration. In the fourth image the outer wings of the distribution are cut via a RF ramp from 150 to 110 kHz. In the following images only a single frequency very close to $\omega_c$, 108.2 and 108.5 kHz, is used in order to cut a hole and a ring into the cloud. The third and sixth image also visualize the second difference between addressing with and without shifting the magnetic trap: the slope grows linearly from the center, which leads to sharper resonances for shifted systems. 

\subsection{Modelling of thermal hopping}

The Arrhenius law is often used to describe chemical reaction rates, but also to model thermal hopping of continuously laser-cooled atoms in very deep optical lattices \cite{Nelson2007S}. Here we use it to model the thermal hopping of ultracold atoms in our two-dimensional lattice. In contrast to quantum mechanical tunneling through the barrier between two lattice sites, thermal hopping refers to motion that is activated thermally when the thermal energy allows to overcome the barrier. To good approximation, the activation energy for a hopping event can be identified with the potential barrier in the lattice, which is  $V_{\rm B}= 9 V_{\rm lat}$ in our triangular lattice convention.

The Arrhenius law describes the hopping rate $\Gamma_{\rm h}$ as the product of an attempt rate $\Gamma_{\rm a}$ and the probability $P(E>V_{\rm B})$ to sample an energy $E$ above the barrier $V_{\rm B}$ in the thermal distribution. 
The hopping rate can then be written as
\begin{equation}
\label{eq:Arrhenius}
\begin{aligned}
    \Gamma_{\rm h} & \approx \Gamma_{\rm a} P(E>V_{\rm B})\\ &
    =\Gamma_{\rm a} \left( \int_{V_{\rm B}}^{\infty}\exp(-E/k_{\rm B} T){\rm d}E \right)/(k_{\rm B} T).
\end{aligned}
\end{equation}

To include quantum tunneling, we add an offset rate, resulting in
\begin{equation}
\label{eq:ArrheniusWithOffset}
\begin{aligned}
    \Gamma_{\rm h} & \approx \Gamma_{\rm a} \exp(-V_{\rm B}/k_B T) + \Gamma_{\rm 0}.
\end{aligned}
\end{equation}

In Fig.~3, we model the temperature-dependent thermalization rate by the modified Arrhenius law of Eq.~(\ref{eq:ArrheniusWithOffset}) and extract an activation barrier of $V_{\rm B}=k_{\rm B} \cdot 2.4(6)\,\mu$K and an attempt rate of $\Gamma_{\rm a}=52(44)\,$Hz as well as an offset rate of $\Gamma_{\rm 0}=0.23(8)\,$Hz, which we attribute to quantum tunneling in higher bands. The barrier height for the calibrated lattice depth of $V_{\rm lat} = 3 E_{\rm r}$ is $V_{\rm B}/k_B = 2.6~\mu$K. We note that in contrast to quantum tunneling, for thermal hopping the atoms can move over long distances in single hopping events. This enables the large scale mass transport in Fig.~3 within few hopping events. 

\subsection{Nanoscale dynamics}

We describe here the numerical simulations shown in Fig.~4C. The simulations start with the ground state of the periodic potential with initial optical lattice beam intensities $I_2,I_3=I_1$. At time $t=0$, $I_2$ and $I_3$ are set to $0.5\cdot I_1$; the intensities change on the intensity lock time scale of about $20\,$µs. For every time step ($5\,$µs) we diagonalize the Hamiltonian of the instantaneous periodic potential and let the state evolve according to the instantaneous eigenstates and eigenvalues. Because the dimers are decoupled from each other, we do not expect any dependence on quasimomentum and we limit ourselves to the $\Gamma$ point in the Brillouin zone. At time $t=390\,$µs, $99.5\%$ of the probability distribution of the time-evolved state is found to lie in the lowest $6$ bands. 

The extracted atomic distribution in a cut of $65\,$nm width is plotted in Fig.~4C (left). In Fig.~4C (middle) the distribution is convoluted with a Gauss filter of $76$nm width, and summed with a offset, for comparison with the experimental data in Fig.~4C (right). The lattice depth used in the theory ($32 E_{\rm rec}$, note that the tunnel barriers are much smaller in a honeycomb lattice compared to a triangular lattice of the same total depth) is calibrated from the comparison with the experiment. The external trap is not included in the analysis, because experimentally we don't see any dependence of the dynamics on the position of the dimer with respect to the trap center.

%%%%%%%%%%%%%%%%%%%%%%%%%%%%%%%%%%%%%%%%%%%%%%%%%%%%%%%%%%%%%%
%%%%%%%%%%%%%%%%%%%%%%%%%%%%%%%%%%%%%%%%%%%%%%%%%%%%%%%%%%%%%%

\section{Discussion of future directions}

\subsection{Prospects for reaching single-atom sensitivity} 

The quantum gas magnifier can be extended to single-atom sensitivity as we demonstrate in the following. This is relevant both for studying arrays of tubes in 3D systems or for imaging single 2D systems, i.e. for reaching strongly-correlated regimes and accessing quantum correlations as in conventional quantum gas microscopes~\cite{Endres2011S}. The estimate below follows the free-space fluorescence imaging of few atoms demonstrated for $^{87}$Rb atoms \cite{Buecker2009S, Fuhrmanek2010S, Picken2017S} and for $^6$Li atoms \cite{Bergschneider2018S}. Alternatively, combining the quantum gas magnifier with the single-atom detection of metastable helium on multi-channel plates \cite{Lawall1995S} would also allow single-atom sensitivity and could even be devised to include a 3D real-space imaging as already achieved in momentum space (see e.g. \cite{Cayla2018S}).

In the following, we exemplarily estimate the expected number of photons from a single $^{87}$Rb atom of mass $m$. The recoil velocity from scattering a photon of wavelength $\lambda=780\,$nm is $v_{\rm rec}=h/(\lambda m) =5.88\,$mm/s. The random recoils lead to a random walk with a width $\sigma(\tau)=\frac{1}{3} v_{\rm rec}R^{1/2}\tau^{3/2}$ after scattering for a time $\tau$ with a rate $R$ \cite{Joffe1993S}. Due to their smaller recoil velocity, heavier atoms allow to scatter more photons before reaching a certain width. For a magnification of $M=70$, the lattice spacing of the triangular lattice $a_{\rm lat}=709\,$nm is magnified to $M a_{\rm lat}=50\,$µm. We therefore want to restrict the width of the random walk to $\sigma=15\,$µm, in order to keep the signal from the different lattice sites separated. This choice restricts the scattering time to $\tau=150\,$µs for a resonant saturated scattering rate $R=\Gamma/2$ with $\Gamma=2\pi \cdot 6\,$MHz, yielding $N=R\tau=2700$ scattered photons. The number of detected photons crucially depends on the numerical aperture (${\rm NA}$), which dictates the solid angle $\Omega=4\pi {\rm NA}^2$, from which the photons are collected. For a medium high ${\rm NA}$ of 0.3, the solid angle covers a fraction $\Omega/(4\pi)=0.09$ of the unit sphere. Assuming a transmission of the imaging system of $80\%$ and a quantum efficiency of the camera of 75$\%$, this yields 145 detected photons. On an Electron Multiplying CCD (EMCCD) camera, an average number of 25 detected photons is sufficient for a signal above the noise level \cite{Bergschneider2018S}. Such a signal can be reached already for scattering for $20\,$µs when a large NA is provided. This estimate does not include the possibly slightly enhanced scattering into the imaging system due to the dipole radiation pattern for correct choice of the magnetic field direction and polarization.

Another limitation to the duration of the optical imaging comes from the fact that the atoms are in free fall. At the end of the time-of-flight expansion of duration $t_{\rm tof}=25\,$ms, they have acquired a velocity $v=g t_{\rm tof}=0.25\,$m/s, where $g=9,81\,$mm/s$^2$ is the gravitational acceleration. To keep the displacement during the imagine pulse of length $t_{\rm im}$ below a lattice constant $M a_{\rm lat}=50\,$µm, it is restricted to $t_{\rm im}\ll M a_{\rm lat}/v=\omega_{\rm ho} a_{\rm lat}/g=200\,$µs. This is no limitation to standard absorption imaging, but becomes relevant for reaching large signals per atoms, in particular when the diffusion from photon scattering allows longer imaging times such as for heavy elements. Solutions include a magnetic levitation during ToF, vertical orientation of the imaging, or the use of fluorescence imaging with carefully imbalanced beam intensities to decelerate the atoms. 

Even without the gravitational acceleration, the atoms obtain a velocity from the matter wave transformation itself, which is given by $v=n a_{\rm lat} \omega_{\rm ho}$ with the initial distance from trap center $n a_{\rm lat}$. This velocity restricts the imaging time to $t_{\rm im}\ll t_{\rm tof}/n$, which could become a limitation for large systems or displacements in the trap with $n\gg 100$. This would still allow larger system sizes than typically achieved in conventional quantum gas microscopes.

\subsection{Prospects for spin-resolved detection}

While we restrict our measurements to atoms in a single spin state, the scheme can be straight-forwardly extended to a spin-resolved detection of spin-mixtures. This is relevant in particular for the Fermi-Hubbard model, but requires complicated protocols in conventional quantum gas microscopes for the detection of both spin states of a single snapshot \cite{Preiss2015S, Boll2016S}.  
One possibility is to choose two spin states with the same magnetic-field dependence, which experience the same harmonic confinement during the $T/4$ pulse. A global microwave sweep or Raman coupling, which transfers one of the states to a state with different magnetic field dependence then allows to spatially separate the two spin states during ToF expansion by adding a magnetic field gradient in a Stern-Gerlach configuration. Using a magnetic-field independent state during the $T/4$ pulse would allow to simultaneously image the real-space distribution for one spin state and the momentum-space distribution for the other spin state. When using a spin-independent optical trap for the $T/4$ pulse, one can work with two spin states with different magnetic moment from the beginning and the step of changing the spin state can be omitted. The compatibility of Stern-Gerlach separation with free-space fluorescence imaging was recently demonstrated in ref.~\cite{Qu2020S}.

\subsection{Local coherence measurements via Talbot interference} \label{section:coherence}

In this section, we discuss how the concept of the quantum gas magnifier -- so far used to image real space density -- could be extended to the realm of coherent phenomena. The Talbot effect in optics describes the revival of a lattice structure with spatial periodicity $L_{\rm Talbot}=2 (M a_{\rm lat})^2/\lambda$ after transmission through a periodic potential with lattice constant $a_{\rm lat}$. The Talbot effect manifest itself also with matter waves \cite{Clauser1994S, Chapman1995S, Nowak1997S} and it is based on the fact that a periodic 1D wavefunction can be decomposed in the plane wave basis at multiples of the wavevector $k=2\pi/a_{\rm lat}$ with kinetic energies $E_n=n^2 (\hbar k)^2/(2m)$ multiple of the same fundamental frequency, whose inverse is the Talbot period $T_{\rm Talbot}$.

We note that in 2D Talbot revivals also appear in the case of triangular and honeycomb lattice because here the allowed wavevectors are ${\bf k}_{n,m}=n {\bf k}_a+m {\bf k}_b$ with ${\bf k}_a=(1,0)$, ${\bf k}_b=(-\frac{1}{2},\frac{\sqrt{3}}{2})$ and $n$, $m$ integers, with associated kinetic energies $E_{n,m}\propto|{\bf k}_{n,m}|^2=n^2+m^2+m\cdot n$. All energies are then integer multiples of the fundamental energy.

The free-space evolution of the density in the traditional Talbot effect can be mapped to the evolution in a HO  by considering the dynamics of the $\tilde{x}$ operator. In free space we have after a time of flight $n T_{\rm Talbot}$ corresponding to the $n^{\rm th}$ Talbot revival:
\begin{equation*}
    \tilde{x}(nT_{\rm Talbot})=\tilde{x}_0+\omega nT_{\rm Talbot}\tilde{p}_0
\end{equation*}
and in the trap: 
\begin{equation*}
    \tilde{x}(t_{\rm ho})=\cos(\omega t_{\rm ho})\tilde{x}_0+\sin(\omega t_{\rm ho})\tilde{p}_0    
\end{equation*}
It follows that for $t_{\rm ho}=\frac{1}{\omega}\arctan(\omega nT_{\rm Talbot})$
\begin{equation*}
    \tilde{x}(t_{\rm ho})=\frac{1}{M'}\tilde{x}(nT_{\rm Talbot})
\end{equation*}
with $M'=\sqrt{1+(\omega nT_{\rm Talbot})^2}=1/\cos(\omega t_{\rm ho})$.\newline
This shows that the dynamics in the trap can be mapped to the dynamics in free-space upon rescaling positions a factor $1/M'$ and 
rescaling of the evolution times via the relation
\begin{equation*} 
    t_{\rm ho}=\frac{1}{\omega}\arctan(\omega nT_{\rm Talbot}).
\end{equation*} 
The distribution in the trap at time $t_{\rm ho}=\frac{1}{\omega}\arctan(\omega nT_{\rm Talbot})$ corresponding to the $n^{\rm th}$ Talbot revival can be then magnified by a factor $M$ just by letting the system remain in the trap for an additional $\sim T/4$ and subsequent $t_{\rm tof}$ expansion. One gets in the end:
\begin{equation*}
    \tilde{x}=\frac{M}{M'}\tilde{x}(nT_{\rm Talbot})\sim \omega t_{\rm tof}\tilde{x}(nT_{\rm Talbot})
\end{equation*}
(since typically $\omega T_{\rm Talbot}\ll 1$ and $\omega t_{\rm tof} \gg 1$)

As derived in \cite{Santra2017S} for the 1D case, the strength of the Talbot revivals is a measure of the phase correlation function. We argue that the decay of the contrast with the order of the revival is related to the phase correlation function also in the 2D case. We note that the quantum gas magnifier would allow access the contrast and therefore the phase correlation function in a spatially resolved manner. This is particularly relevant for inhomogeneous systems, as typically the case for harmonically trapped quantum gases.

Imaging of coherent wavepackets away from the focusing condition of the quantum gas magnifier can be used to gain information on phase profiles of the wavefunction, where phase fluctuations of low-dimensional systems are transformed into density fluctuations \cite{Dettmer2001S, Choi2012S}. As an example, we suggest it could be used to detect domains of magnetic order encoded in the condensate phases \cite{Struck2011S}. 

These measurement would take place by adding an evolution time in the HO, followed by the magnification whose first step is also an evolution in the HO itself: we notice that one could also measure negative waiting times simply by a total wait time in the HO smaller than required from the precise imaging of the density $\omega t_{\rm ho}<\arctan(\omega t_{\rm tof})$. Measuring for both positive and negative times would allow detection of time reversal asymmetry, a strong hint for chiral states.

%\bibliography{mybib}


%\bibliographystyle{apsrev4-1}
%\bibliographystyle{naturemag}


\setcounter{equation}{0}
\setcounter{figure}{0}
\setcounter{table}{0}
%\setcounter{page}{1}
%\makeatletter
\renewcommand{\theequation}{S\arabic{equation}}
\renewcommand{\thefigure}{S\arabic{figure}}
\renewcommand{\bibnumfmt}[1]{[S#1]}
\renewcommand{\citenumfont}[1]{S#1}

%%%%%%%%%%%%%%%%%%%%%%%%%%%%%%%%%%%%%%%%%%%%%%%%%%%%%%%%%%%%%%5